\documentclass[twocolumn]{aastex631}

\def\afrhot{{$A$($\theta$)$f\rho$}}
\def\afrhoz{{$A$(0$^\circ$)$f\rho$}}
\def\afrho{{$Af\rho$}}
\def\degr{$^\circ$}
\def\rh{$r_{\mathrm{H}}$}

\received{December 24, 2020}
\accepted{March 16, 2021}

\submitjournal{the A'Hearn Symposium special issue of {\it The Planetary Science Journal}}

\shorttitle{Narrowband Observations of Comet Wirtanen II}
\shortauthors{Knight et al.}
\graphicspath{{./}{figures/}}

\begin{document}

\title{Narrowband Observations of Comet 46P/Wirtanen During its Exceptional Apparition of 2018/19 II: Photometry, Jet Morphology, and Modeling Results}

\email{knight@usna.edu}

\author[0000-0003-2781-6897]{Matthew M. Knight}
\affiliation{United States Naval Academy, Department of Physics, 572C Holloway Rd, Annapolis, MD 21402, USA}
\affiliation{University of Maryland, Department of Astronomy, College Park, MD 20742, USA}

\author[0000-0003-1412-2511]{David G. Schleicher}
\affiliation{Lowell Observatory, 1400 W. Mars Hill Rd, Flagstaff, AZ 86001, USA}

\author[0000-0002-4767-9861]{Tony L. Farnham}
\affiliation{University of Maryland, Department of Astronomy, College Park, MD 20742, USA}



\begin{abstract}

We report on our extensive photometry and imaging of Comet 46P/Wirtanen during its 2018/19 apparition and use these data to constrain modeling of Wirtanen's activity. Narrowband photometry was obtained on nine epochs from 2018 October through 2019 March as well as 10 epochs during the 1991, 1997, and 2008 apparitions. The ensemble photometry reveals a typical composition and a secular decrease in activity since 1991. Production rates were roughly symmetric around perihelion for the carbon-bearing species (CN, C$_3$, and C$_2$), but steeper for OH and NH outbound. Our imaging program emphasized CN, whose coma morphology and lightcurve yielded rotation periods reported in a companion paper (Farnham et al., PSJ, 2, 7). Here, we compare the gas and dust morphology on the 18 nights for which observations of additional species were obtained. The carbon-bearing species exhibited similar morphology that varied with rotation. OH and NH had broad, hemispheric brightness enhancements in the tailward direction that did not change significantly with rotation, which we attribute to their originating from a substantial icy grain component.
We constructed a Monte Carlo model that replicates the shape, motion, and brightness distribution of the CN coma throughout the apparition with a single, self-consistent solution in principal axis rotation. Our model yields a pole having (R.A., Decl.) =  319$^\circ$, $-$5$^\circ$ (pole obliquity of 70$^\circ$) and two large sources (radii of 50$^\circ$ and 40$^\circ$) centered at near-equatorial latitudes and separated in longitude by $\sim$160$^\circ$. 
Applications of the model to explain observed behaviors are discussed.

\end{abstract}



\section{Introduction} \label{sec:intro}

Comet 46P/Wirtanen was discovered in 1948 by Carl Wirtanen at Lick Observatory \citep{jeffers48}. Despite being observed at every perihelion since then except 1980 \citep[e.g.,][]{kronk09,kronk10,kronk17}, little was known about it prior to its selection as the target of the {\it Rosetta} mission in 1994. Despite a relatively unfavorable apparition in 1997, many investigations were undertaken in order to characterize Wirtanen before {\it Rosetta} launched, notably including the measurement of its small nucleus size \citep[radius $\sim$0.6~km;][]{lamy98} 
and thus very high active fraction \citep{farnham98}. 
Ultimately, delays in {\it Rosetta}'s launch caused Wirtanen to be scrapped in favor of comet 67P/Churyumov-Gerasimenko in 2004 \citep{ulamec06}. Wirtanen's favorable orbit has continued to make it a compelling mission target, leading to its selection for at least one subsequent mission proposal -- Comet Hopper, a finalist for a NASA Discovery mission in 2011.

In the last century, Wirtanen's orbit has undergone a series of perturbations from Jupiter that have resulted in both smaller perihelion distances and shorter orbital periods. The most recent significant perturbation, in 1984, reduced its perihelion distance to 1.08 AU. As a result, the 2018/19 apparition was its most favorable on record, and the close approach distance of 0.0775 AU is the 14$^\mathrm{th}$ closest approach of a comet to Earth in the last 200 years\footnote{{\tt http://wirtanen.astro.umd.edu/close\_approaches.shtml}}. Wirtanen was more active than other recent close approaching comets including 209P/LINEAR, 45P/Honda-Mrkos-Pajdu\v s\'akov\'a, and 41P/Tuttle-Giacobini-Kres\'ak. As a result, Wirtanen's 2018/19 apparition was one of the best observing opportunities for a Jupiter family comet in many years, particularly for northern hemisphere observers, and it was the target of a widespread observation campaign.  Results published so far include measurements of the nucleus' rotation period \citep{farnham21}, detections of at least five small ($<$1.5 mag) outbursts and discovery of a dust trail \citep{farnham19,farnham21,kelley19,kelley20}, the first measurement of the D/H ratio \citep{lis19}, and detection of HCN \citep{wang20}. 

We planned an extensive campaign using the telescopes at Lowell Observatory to conduct photoelectric photometry and imaging at near-UV and optical wavelengths. 
Our objectives during the 2018/19 apparition were numerous and designed to capitalize on the rare opportunity to observe the comet while it was bright for several months over a wide range of viewing geometries and at high spatial resolution. These included obtaining extensive photoelectric photometry for comparison to previous datasets to look for secular evolution and better quantify seasonal behaviors, identifying diagnostic features in the gas and dust coma, and using the motion and repetition of these features to constrain the nucleus' rotation period at one or more epochs. Ultimately, we collected usable data on more than 40 nights and were able to construct a detailed model of the nucleus including the location and extent of source regions and orientation of the rotation pole.

Results of our extensive CN imaging, including identification of at least two gas jets and evidence for the nucleus being in simple rotation with an apparent period that increased from 8.98 hr to 9.14 hr, then decreased to 8.94 hr, are discussed in a companion paper, \citet{farnham21} and henceforth referred to as Paper I. The current paper presents complementary analyses, some of which rely on the results from Paper I. Observations and reductions are summarized in Section~\ref{sec:observations}. In Section~\ref{sec:photometry} we analyze photometry from this and three previous apparitions. In Section~\ref{sec:imaging} we discuss the coma morphology of dust and gas species other than CN. The photometry and imaging are used as constraints for our Monte Carlo model in Section~\ref{sec:modeling}. The results and conclusions are discussed in Sections~\ref{sec:discussion} and \ref{sec:conclusion}, respectively.

\section{OBSERVATIONS AND REDUCTIONS}
\label{sec:observations}

\subsection{CCD Observations and Reductions}
\label{sec:obs_ccd}
As discussed in Paper I, we imaged Wirtanen extensively throughout the 2018/19 apparition using the telescopes at Lowell Observatory. We made the first successful recovery of the apparition in June 2018 \citep{mpec2018n81} and monitored it thereafter, but it was too faint for the studies reported here until October. Regular monitoring was conducted from late October 2018 through early February 2019 using the 31in (0.8m) telescope in robotic mode, while dedicated observing runs were made using the John S.\ Hall 42in (1.1m) and the 4.3-m Lowell Discovery Telescope (LDT; then called Discovery Channel Telescope or DCT) in early-November, early- and mid-December, and early- and late-January. Most 31in nights consisted of ``snapshot'' (one epoch of observations) or ``monitoring'' (two or more epochs of non-continuous observations) observations using broadband {\it r}$'$ or {\it R} filters and a narrowband CN filter from the HB comet filter set \citep{farnham00}. Wirtanen was imaged nearly continuously on most 42in and LDT nights, with regular observations in {\it R} or {\it r$'$} and CN supplemented with other HB narrowband filters designed to isolate gas (OH, NH, C$_2$, C$_3$) or dust (UC, GC, BC, RC). A complete overview of observations is given in Table~\ref{t:imaging_circ}.

\renewcommand{\baselinestretch}{0.6} 
\renewcommand{\arraystretch}{1.2}

\begin{deluxetable*}{lccrccccccc}  
\tabletypesize{\scriptsize}
\tablecolumns{11}
\tablewidth{0pt} 
\setlength{\tabcolsep}{0.05in}
\tablecaption{Imaging observations and geometric parameters for comet 46P/Wirtanen.\tablenotemark{\scriptsize{a}}}
\tablehead{   
  \colhead{UT}&
  \colhead{UT}&
  \colhead{Tel.\tablenotemark{\scriptsize{b}}}&
  \colhead{$\Delta$T\tablenotemark{\scriptsize{c}}}&
  \colhead{$r_\mathrm{H}$\tablenotemark{\scriptsize{d}}}&
  \colhead{$\Delta$\tablenotemark{\scriptsize{e}}}&
  \colhead{$\alpha$\tablenotemark{\scriptsize{f}}}&
  \colhead{P.A.\tablenotemark{\scriptsize{g}}}&
  \colhead{Filters}&
  \colhead{Obs.}&
  \colhead{Cond.\tablenotemark{\scriptsize{i}}} \\
  \colhead{Date}&
  \colhead{Range}&
  \colhead{}&
  \colhead{(day)}&
  \colhead{(AU)}&
  \colhead{(AU)}&
  \colhead{($^\circ$)}&
  \colhead{($^\circ$)}&
  \colhead{}&
  \colhead{Type\tablenotemark{\scriptsize{h}}}&
  \colhead{}
}
\startdata
2018 Oct 28&7:22$-$\phantom{0}7:36&31in&$-$45.62&1.218&0.295&35.8&183.4&{\it R}, CN&S&CLR\\
2018 Nov 1&4:37$-$\phantom{0}8:38&42in&$-$41.65&1.193&0.273&38.0&188.1&{\it R}, CN, OH, C3, C2, BC, GC&C&CIR\\
2018 Nov 1&5:52$-$\phantom{0}7:29&31in&$-$41.65&1.193&0.273&38.0&188.1&{\it R}, CN&M&CIR\\
2018 Nov 2&4:30$-$\phantom{0}8:55&42in&$-$40.65&1.187&0.267&38.6&189.2&{\it R}, CN, OH, C3, C2, BC, GC&C&CIR\\
2018 Nov 3&4:21$-$\phantom{0}8:53&LDT&$-$39.65&1.181&0.262&39.1&190.3&{\it r}$'$, CN&C&CLD\\
2018 Nov 4&4:13$-$\phantom{0}8:34&42in&$-$38.66&1.176&0.256&39.7&191.4&{\it R}, CN, OH, C3, C2, BC, GC&C&CIR\\
2018 Nov 4&5:40$-$\phantom{0}7:25&31in&$-$38.65&1.176&0.256&39.7&191.4&{\it R}, CN&M&CIR\\
2018 Nov 9&4:36$-$\phantom{0}7:48&42in&$-$33.67&1.148&0.229&42.3&196.6&{\it R}, CN&C&CLR\\
2018 Nov 10&4:36$-$\phantom{0}5:04&31in&$-$32.73&1.143&0.224&42.7&197.5&{\it R}, CN&S&CLR\\
2018 Nov 11&4:36$-$\phantom{0}5:03&31in&$-$31.73&1.138&0.219&43.2&198.5&{\it R}, CN&S&CLR*\\
2018 Nov 11&4:16$-$\phantom{0}8:01&42in&$-$31.67&1.138&0.219&43.2&198.6&{\it R}, CN&C&CLR*\\
2018 Nov 12&4:36$-$\phantom{0}5:04&31in&$-$30.73&1.134&0.214&43.7&199.5&{\it R}, CN&S&CLR\\
2018 Nov 12&4:15$-$\phantom{0}7:26&42in&$-$30.68&1.133&0.214&43.7&199.5&{\it R}, CN&C&CLR\\
2018 Nov 13&5:07$-$\phantom{0}5:27&31in&$-$29.71&1.129&0.208&44.1&200.4&{\it R}, CN&S&CLR\\
2018 Nov 13&4:34$-$\phantom{0}7:45&42in&$-$29.67&1.128&0.208&44.1&200.5&{\it R}, CN&C&CLR\\
2018 Nov 15&5:00$-$\phantom{0}5:20&31in&$-$27.71&1.120&0.198&44.9&202.2&{\it R}, CN&S&CLR\\
2018 Nov 16&4:58$-$\phantom{0}5:18&31in&$-$26.71&1.115&0.193&45.3&203.1&{\it R}, CN&S&CLD\\
2018 Nov 26&4:33$-$\phantom{0}6:16&31in&$-$16.70&1.079&0.143&46.7&211.3&{\it R}, CN&S&CLD\\
2018 Nov 27&4:29$-$\phantom{0}6:14&31in&$-$15.70&1.077&0.138&46.5&212.1&{\it R}, CN&S&CLR\\
2018 Nov 29&5:27$-$\phantom{0}6:10&31in&$-$13.69&1.072&0.128&45.8&213.7&{\it R}, CN&S&CLD\\
2018 Dec 3&2:06$-$\phantom{0}9:05&LDT&$-$9.69&1.064&0.111&43.3&217.3&{\it r}$'$, CN, OH, NH, C3, C2, UC, BC, RC&C&ICL\\
2018 Dec 4&1:55$-$\phantom{0}8:41&LDT&$-$8.71&1.062&0.107&42.3&218.3&{\it r}$'$, CN, OH, C3, C2, BC, RC&C&ICL\\
2018 Dec 5&6:05$-$\phantom{0}6:23&31in&$-$7.67&1.060&0.103&41.1&219.6&{\it R}, CN&S&CLR\\
2018 Dec 6&1:36$-$\phantom{0}6:57&LDT&$-$6.75&1.059&0.099&39.9&220.7&{\it R}, CN, OH&M&CLD\\
2018 Dec 9&1:46$-$\phantom{0}2:07&42in&$-$3.85&1.057&0.089&35.2&225.7&{\it R}, CN&S&CLR\\
2018 Dec 10&0:46$-$10:11&42in&$-$2.70&1.056&0.086&33.0&228.5&{\it R}, CN, OH, NH, C3, C2&C&CIR\\
2018 Dec 12&1:47$-$\phantom{0}7:10&LDT&$-$0.74&1.055&0.082&28.9&234.9&{\it R}, CN, OH, NH, C3, BC&C&CIR\\
2018 Dec 12&2:06$-$\phantom{0}9:03&31in&$-$0.70&1.055&0.082&28.8&235.0&{\it R}, CN&M&CIR\\
2018 Dec 13&1:15$-$\phantom{0}7:25&LDT&$+$0.25&1.055&0.080&26.7&239.2&{\it R}, CN, OH, NH, C3, BC&C&CLR\\
2018 Dec 14&1:08$-$\phantom{0}7:15&LDT&$+$1.25&1.055&0.079&24.5&244.6&{\it R}, CN, OH, NH, C3, BC&C&CLR\\
2018 Dec 15&1:32$-$\phantom{0}9:49&LDT&$+$2.31&1.056&0.078&22.3&251.7&{\it r}$'$, CN&C&CLD\\
2018 Dec 16&1:40$-$11:26&LDT&$+$3.35&1.056&0.077&20.5&260.2&{\it r}$'$, CN, OH, NH, C3, C2, UC, BC, RC&C&CLR\\
2018 Dec 17&1:30$-$11:18&LDT&$+$4.34&1.057&0.078&19.2&239.7&{\it r}$'$, CN, OH, NH, C3, C2, UC, BC, RC&C&CLD\\
2018 Dec 19&2:15$-$11:23&31in&$+$6.36&1.059&0.079&18.1&291.9&{\it R}, CN&M&CLR\\
2018 Dec 23&2:23$-$\phantom{0}2:41&31in&$+$10.18&1.064&0.087&21.1&328.5&{\it R}, CN&S&CLD\\
2018 Dec 24&5:24$-$\phantom{0}5:43&31in&$+$11.30&1.066&0.090&22.6&336.3&{\it R}, CN&S&CLD\\
2018 Dec 25&3:25$-$12:34&31in&$+$12.40&1.069&0.094&24.1&342.7&{\it R}, CN&M&ICL\\
2018 Dec 27&3:22$-$13:06&31in&$+$14.41&1.073&0.102&26.6&352.5&{\it R}, CN&M&CLD\\
2018 Dec 30&2:27$-$13:24&31in&$+$17.40&1.081&0.115&29.6&\phantom{00}3.5&{\it R}, CN&M&CIR\\
2018 Dec 31&2:29$-$\phantom{0}4:04&31in&$+$18.26&1.084&0.119&30.2&\phantom{00}6.1&{\it R}, CN&S&CLD\\
2019 Jan 3&1:45$-$13:12&42in&$+$21.38&1.094&0.134&32.1&\phantom{0}13.7&{\it R}, CN, OH&C&CLR\\
2019 Jan 4&4:06$-$13:31&31in&$+$22.44&1.098&0.139&32.5&\phantom{0}15.7&{\it R}, CN&S&CLR\\
2019 Jan 4&4:14$-$13:12&42in&$+$22.43&1.098&0.139&32.5&\phantom{0}15.6&{\it R}, CN, OH&C&CLR\\
2019 Jan 5&3:42$-$13:30&31in&$+$23.43&1.102&0.145&32.8&\phantom{0}17.3&{\it R}, CN&M&ICL\\
2019 Jan 12&1:43$-$13:40&LDT&$+$30.39&1.132&0.184&33.4&\phantom{0}23.4&{\it r}$'$, CN, OH, BC, RC&C&ICL\\
2019 Jan 26&2:01$-$13:12&42in&$+$44.39&1.210&0.273&30.5&\phantom{0}17.4&{\it R}, CN, OH&C&ICL\\
2019 Jan 27&2:01$-$\phantom{0}6:32&42in&$+$45.25&1.215&0.279&30.3&\phantom{0}16.6&{\it R}, CN, OH&C&ICL\\
2019 Jan 28&2:00$-$13:06&42in&$+$46.39&1.223&0.287&30.0&\phantom{0}15.4&{\it R}, CN, OH&C&CLR\\
2019 Feb 8&4:10$-$\phantom{0}4:40&31in&$+$57.26&1.298&0.368&27.6&\phantom{00}1.9&{\it R}, CN&S&CLR\\
2019 Feb 8&4:00$-$13:29&42in&$+$57.44&1.299&0.369&27.6&\phantom{00}1.6&{\it R}, CN&C&CLR\\
2019 Feb 8&2:16$-$12:25&LDT&$+$57.38&1.298&0.369&27.6&\phantom{00}1.7&{\it R}, CN&M&CIR\\
2019 Feb 9&2:33$-$\phantom{0}5:34&42in&$+$58.24&1.305&0.376&27.4&\phantom{00}0.6&{\it R}, CN&C&CLD\\
\enddata
\tablenotetext{a} {All parameters are given for the midpoint of each night's observations.}
\tablenotetext{b} {Telescope used: LDT = Lowell Discovery Telescope (4.3-m), 42in = Hall 42-in Telescope (1.1-m), 31in = 31-in Telescope (0.8-m).}
\tablenotetext{c} {Time from perihelion.}
\tablenotetext{d} {Heliocentric distance.}
\tablenotetext{e} {Geocentric distance.}
\tablenotetext{f} {Solar phase angle.}
\tablenotetext{g} {Position angle of the Sun.}
\tablenotetext{h} {Type of observation: S = snapshot (one epoch during the night), M = monitoring (2 or more visits during a night with gaps), C = continuous (nearly continuous observations throughout the night)}
\tablenotetext{i} {Conditions: CLR = clear, CIR = cirrus, ICL = intermittent clouds, CLD = clouds; * indicates possible smoke}
\label{t:imaging_circ}
\end{deluxetable*}

\vspace{-8.5mm}
LDT images used the Large Monolithic Imager (LMI) which has a 6.1 K ${\times}$ 6.1 K e2v CCD with a field of view of 12.3 arcmin on a side. Depending on conditions and the observational goals for a night, images were binned on chip to either $2{\times}2$ or $3{\times}3$ resulting in pixel scales of 0.24 arcsec and 0.36 arcsec, respectively. The 42in's NASA42 camera has an e2v CCD231-84 chip with 4 K $\times$ 4 K pixels and a field of view 25.3 arcmin on a side. On chip binning of $2{\times}2$ or $3{\times}3$ resulted in pixel scales of 0.74 arcsec and 1.11 arcsec, respectively. The 31in has a 2 K $\times$ 2 K e2v CCD42-40 with 0.46 arcsec pixels and a field of view about 15.7 arcmin on a side. 

All images were acquired at the comet's ephemeris rate. Exposure times varied with the comet's brightness, telescope size, and time available, but were typically 30--60 sec for broadband images and 180--300 sec for narrowband images. Most filters were acquired in sets of 3--5 images which were later median combined to improve signal-to-noise and minimize the effects of background stars, bad pixels, cosmic rays, etc. When conditions were photometric during our dedicated observing runs, HB standard stars \citep{farnham00} were obtained to allow absolute calibrations of narrowband images. 

Images were debiased and flat-field corrected using standard procedures. Absolute calibrations of narrowband images were performed on photometric nights following the procedures outlined in \citet{farnham00} to remove underlying continuum from gas images (OH, NH, CN, C$_2$, and C$_3$). 
The standard reduction methodology assumes a solar dust color if a particular continuum filter is not available; we found that this yielded an over-subtraction of continuum in the NH filter if the UV continuum (UC) filter was not used. Thus, we processed December 13 and 14 by assuming the UC to blue continuum (BC) ratio was the same as on December 16. 
We confirmed that this technique yielded comparable morphology to the standard procedure on other nights where UC and BC were both obtained, so we are confident that it could safely be used to assess morphology on these nights. 
We confirmed on the photometric nights that the CN and OH were minimally contaminated with dust and that the bulk morphology was unaffected by the continuum removal. Thus, we also used CN and OH for morphological studies on non-photometric nights. 

We determined centroids using a 2D Gaussian fit to the inner coma and used these centroids for photometric measurements, aligning images, and for applying our standard image enhancement routines \citep[e.g.,][]{schleicher04,knight17a}. A variety of image enhancement techniques were utilized when exploring the data, with removal of a temporal average (based on the period determined in Paper I) preferred when sufficient data were available, and removal of an azimuthal median favored otherwise. Both enhancements are relatively benign and do not introduce or significantly alter structures in the coma, although we do use caution in interpreting features within a few pixels of the center as these are most sensitive to the centroiding. 
When needed, Gaussian smoothing was applied to images having lower signal-to-noise ratios.

\subsection{Photometer Observations and Reductions}
\label{sec:obs_phot}
Narrowband photometry was obtained with two traditional photoelectric photometers at the 
John S. Hall 42-inch (1.1-m) telescope at Lowell Observatory, except for two nights in 
1997 when the 31-inch (0.8-m) telescope was used instead. The IHW filter set \citep{osborn90,ahearn91,larson91} was utilized 
in 1991 and for all but one set of data in 1997, while the HB filter set was used in that 
one case and for all observations during the 2007/08 and 2018/19 apparitions. 

Standard observing and reduction techniques and coefficients were employed for the 
photometric observations, as detailed by \citet{farnham00} for the HB filters and 
as revised by \citet{farnham05} for the older IHW filters. Because these 
revisions for the IHW filters gave improved decontamination for the continuum filters, 
for completeness we re-tabulate here the adjusted values for the originally published 
results from \citet{farnham98} along with the more recent apparitions. In all cases, fluxes 
within the entrance aperture of the photometer are converted to column abundances, $M(\rho)$, 
and then to production rates, $Q$, for each of the five observed gas species -- OH, NH, 
CN, C$_2$, and C$_3$ \citep[cf.][]{ahearn95}. Continuum fluxes (note that wavelengths 
of the continuum filters changed between the two filter sets) are converted to 
$A({\theta})f{\rho}$, a proxy for dust production introduced by \citet{ahearn84}. 
Phase angle adjustments are made using the Schleicher-Marcus composite dust phase curve   
\citep[see][]{schleicher11}.

\vspace{3mm}
\section{PHOTOMETRIC RESULTS}
\label{sec:photometry}

\vspace{2mm}
\subsection{The Photometry Datasets}
Since rotational studies during the 2018/19 apparition
would best be conducted using imaging techniques, our goals for the photometry were 
primarily intended to extend our heliocentric distance coverage of Wirtanen as compared 
to what was obtained at prior apparitions and to search for possible secular trends. 
We, therefore, employed the methodology of trying to acquire one photometric night of 
data per observing run, with subsequent nights used for additional imaging. 

\begin{figure*}
  \centering
  \includegraphics[width=160mm]{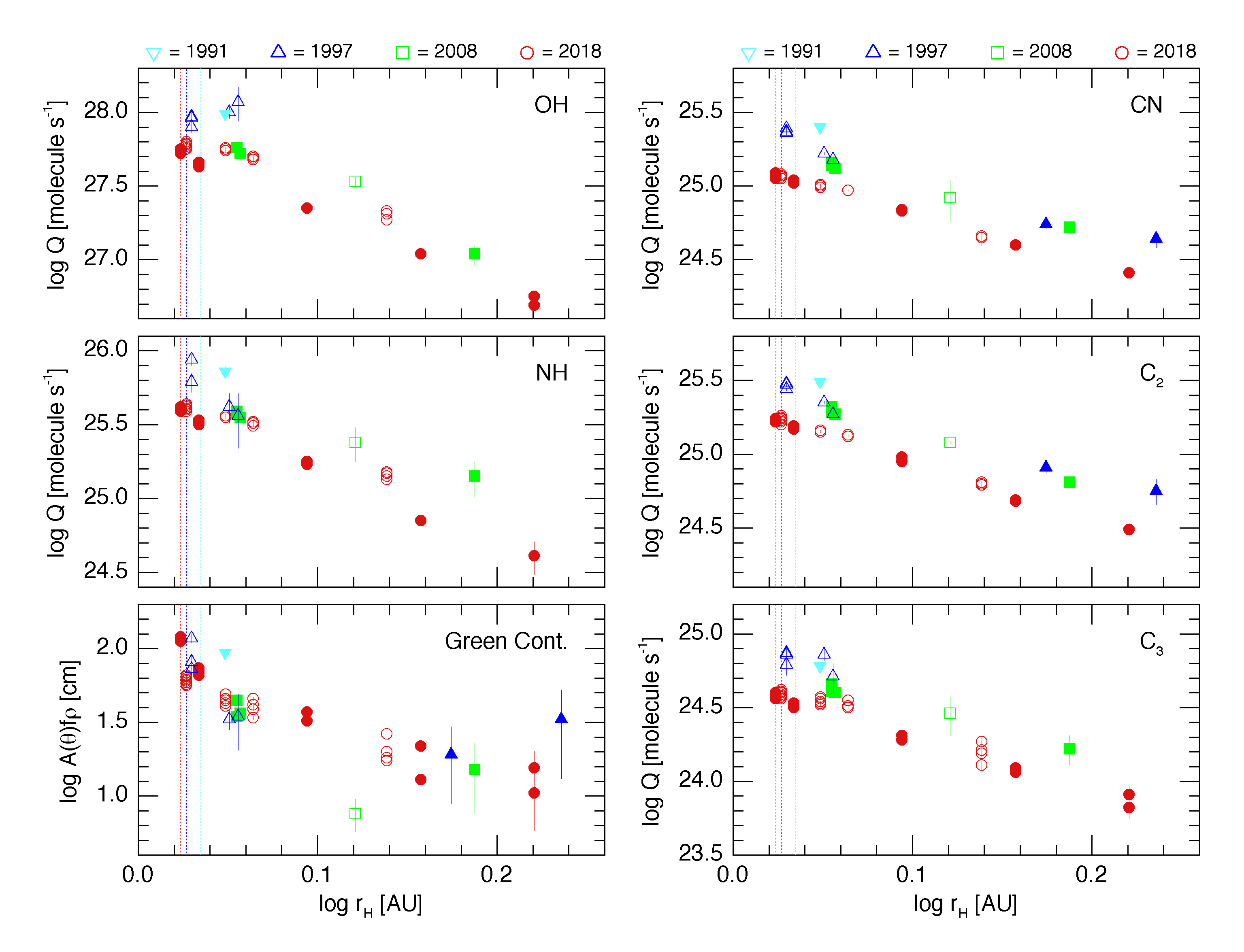}
  \caption{Logarithm of the production rates for observed molecular species and $A(\theta)f\rho$ for the dust as a function of the logarithm of the heliocentric distance. The species is indicated in the upper right of each plot. Different symbols are used for each apparition and are given above the plots; open points are pre-perihelion and filled points are post-perihelion. Error bars are plotted for all points; in some cases they are smaller than the symbols and are therefore not visible. Aperture sizes and other relevant parameters are given in Tables~\ref{t:phot_circ}--\ref{t:phot_rates}.}
  \label{fig:photometry}
\end{figure*}

This method worked quite well overall, with only 
two runs -- in late September and late October -- completely clouded out, while nine 
other runs were successful. Observing circumstances for these nine nights, along with 
four nights from 2007/08, five nights from 1997, and one from 1991, are listed in 
Table~\ref{t:phot_circ}. While most nights in the first three apparitions have only a single dataset, 
all nights in 2018/19 have at least two sets, usually with multiple aperture sizes, 
and on one night we obtained 12 sets using 9 different apertures. Reduced fluxes and 
resulting aperture abundances are given in Table~\ref{t:phot_flux}, while the derived gas production 
rates and dust \afrho\ values are listed in Table~\ref{t:phot_rates}. These final results are also 
plotted, as logarithms, as a function of log \rh\ in Figure~\ref{fig:photometry}. Different symbols 
distinguish each apparition while pre- and post-perihelion data are identified by 
open and filled symbols, respectively. Since the photometric uncertainties (1-$\sigma$)
are unbalanced in logarithmic space, only the ``positive'' uncertainties 
are tabulated, while the ``negative'' uncertainties can be readily calculated.

\renewcommand{\baselinestretch}{0.75} 
\renewcommand{\arraystretch}{1.2}

\begin{deluxetable*}{llllcccccccc}  
\tabletypesize{\scriptsize}
\tablecolumns{12}
\tablewidth{0pt} 
\setlength{\tabcolsep}{0.03in}
\tablecaption{Photometry observing circumstances and fluorescence efficiencies for comet 46P/Wirtanen.\tablenotemark{\scriptsize{a}}}
\tablehead{   
  \multicolumn{3}{c}{UT Date}&
  \colhead{$\Delta$T}&
  \colhead{$r_\mathrm{H}$}&
  \colhead{$\Delta$}&
  \colhead{Phase}&
  \colhead{Phase}&
  \colhead{$\dot{r}_\mathrm{H}$}&
  \multicolumn{3}{c}{log $L/N$ (erg s$^{-1}$ molecule$^{-1}$)} \\
  \cline{10-12}
  \colhead{}&
  \colhead{}&
  \colhead{}&
  \colhead{(day)}&
  \colhead{(AU)}&
  \colhead{(AU)}&
  \colhead{($^\circ$)}&
  \colhead{Adj.\tablenotemark{\scriptsize{b}}}&
  \colhead{(km s$^{-1}$)}&
  \colhead{OH}&
  \colhead{NH}&
  \colhead{CN}
}
\startdata
\vspace{0.05in}
1991&Oct& 11.5&\phantom{0}+21.0&1.118&1.408&\phantom{0}44.8&0.457&\phantom{0}+5.6&$-$14.674&$-$13.157&$-$12.424\\
1997&Feb& 12.1&\phantom{0}$-$30.4&1.137&1.600&\phantom{0}37.7&0.429&\phantom{0}$-$8.1&$-$14.876&$-$13.226&$-$12.526\\
1997&Feb& 15.1&\phantom{0}$-$27.4&1.124&1.590&\phantom{0}38.0&0.431&\phantom{0}$-$7.4&$-$14.886&$-$13.215&$-$12.520\\
1997&Mar&\phantom{0}5.1&\phantom{0}\phantom{0}$-$9.4&1.071&1.536&\phantom{0}39.9&0.439&\phantom{0}$-$2.7&$-$14.833&$-$13.251&$-$12.573\\
1997&Jun&\phantom{0}4.2&\phantom{0}+81.7&1.494&1.961&\phantom{0}30.5&0.386&+14.2&$-$14.585&$-$13.469&$-$12.724\\
\vspace{0.05in}
1997&Jul&\phantom{0}1.2&+108.7&1.721&2.311&\phantom{0}24.0&0.331&+14.7&$-$14.699&$-$13.595&$-$12.845\\
2007&Dec&\phantom{0}4.2&\phantom{0}$-$60.3&1.321&1.080&\phantom{0}47.2&0.464&$-$12.8&$-$14.790&$-$13.379&$-$12.706\\
2008&Mar&\phantom{0}4.2&\phantom{0}+30.7&1.135&0.941&\phantom{0}56.1&0.469&\phantom{0}+8.3&$-$14.672&$-$13.158&$-$12.432\\
2008&Mar&\phantom{0}5.2&\phantom{0}+31.7&1.140&0.944&\phantom{0}55.9&0.469&\phantom{0}+8.5&$-$14.674&$-$13.164&$-$12.435\\
\vspace{0.05in}
2008&Apr& 30.2&\phantom{0}+87.8&1.540&1.435&\phantom{0}39.4&0.437&+14.5&$-$14.606&$-$13.498&$-$12.747\\
2018&Oct&\phantom{0}6.3&\phantom{0}$-$67.6&1.376&0.438&\phantom{0}25.8&0.348&$-$13.5&$-$14.860&$-$13.415&$-$12.738\\
2018&Nov&\phantom{0}7.3&\phantom{0}$-$35.7&1.159&0.240&\phantom{0}41.3&0.445&\phantom{0}$-$9.4&$-$14.790&$-$13.253&$-$12.550\\
2018&Nov& 15.3&\phantom{0}$-$27.6&1.119&0.198&\phantom{0}44.9&0.458&\phantom{0}$-$7.7&$-$14.876&$-$13.212&$-$12.514\\
2018&Dec&\phantom{0}3.2&\phantom{0}\phantom{0}$-$9.8&1.064&0.111&\phantom{0}43.3&0.452&\phantom{0}$-$2.9&$-$14.824&$-$13.240&$-$12.561\\
2018&Dec& 16.3&\phantom{0}\phantom{0}+3.4&1.056&0.077&\phantom{0}20.5&0.296&\phantom{0}+1.0&$-$14.854&$-$13.220&$-$12.593\\
2018&Dec& 30.1&\phantom{0}+17.2&1.081&0.113&\phantom{0}29.4&0.378&\phantom{0}+5.0&$-$14.660&$-$13.138&$-$12.403\\
2019&Jan& 31.2&\phantom{0}+49.3&1.242&0.307&\phantom{0}29.3&0.377&+11.6&$-$14.569&$-$13.278&$-$12.535\\
2019&Feb& 26.2&\phantom{0}+75.3&1.437&0.529&\phantom{0}26.2&0.352&+14.0&$-$14.556&$-$13.434&$-$12.690\\
2019&Mar& 25.2&+102.2&1.662&0.843&\phantom{0}27.8&0.365&+14.8&$-$14.668&$-$13.567&$-$12.812\\
\enddata
\tablenotetext{a} {All parameters are given for the midpoint of each night's observations.}
\tablenotetext{b} {Adjustment to 0$^\circ$ solar phase angle to log({\afrhot}) values based on adopted phase function (see text).}
\label{t:phot_circ}
\end{deluxetable*}

\begin{deluxetable*}{lcccc@{\extracolsep{4pt}}ccc@{\extracolsep{4pt}}ccc@{\extracolsep{4pt}}ccccccc}  
\tabletypesize{\scriptsize}
\tablecolumns{18}
\tablewidth{0pt} 
\setlength{\tabcolsep}{0.02in}
\tablecaption{Photometric fluxes and aperture abundances for comet 46P/Wirtanen.}
\tablehead{   
  \multicolumn{3}{c}{UT Date}&
  \multicolumn{2}{c}{Aperture}&
  \multicolumn{5}{c}{log Emission Band Flux}&
  \multicolumn{3}{c}{log Continuum Flux\tablenotemark{\scriptsize{a}}}&
  \multicolumn{5}{c}{log $M$($\rho$)} \\
  \cline{4-5}
  \colhead{}&
  \colhead{}&
  \colhead{}&
  \colhead{Size}&
  \colhead{log $\rho$}&
  \multicolumn{5}{c}{(erg cm$^{-2}$ s$^{-1}$)}&
  \multicolumn{3}{c}{(erg cm$^{-2}$ s$^{-1}$ \AA$^{-1}$)}&
  \multicolumn{5}{c}{(molecule)} \\
  \cline{6-10}
  \cline{11-13}
  \cline{14-18}
  \colhead{}&
  \colhead{}&
  \colhead{}&
  \colhead{(arcsec)}&
  \colhead{(km)}&
  \colhead{OH}&
  \colhead{NH}&
  \colhead{CN}&
  \colhead{C$_3$}&
  \colhead{C$_2$}&
  \colhead{UV}&
  \colhead{Blue}&
  \colhead{Green}&
  \colhead{OH}&
  \colhead{NH}&
  \colhead{CN}&
  \colhead{C$_3$}&
  \colhead{C$_2$}
}
\startdata
\vspace{0.05in}
1991&Oct&11.5&\phantom{0}35.3&4.26&$-$10.43&$-$11.27&$-$10.62&$-$10.87&$-$10.74&$-$14.16&...&$-$13.84&31.99&29.64&29.55&28.97&29.45\\
1997&Feb&12.1&\phantom{0}73.7&4.63&$-$10.15&$-$11.19&$-$10.58&$-$10.83&$-$10.61&$-$14.13&...&$-$14.01&32.58&29.89&29.80&29.14&29.70\\
1997&Feb&15.1&\phantom{0}73.7&4.63&$-$10.23&$-$11.11&$-$10.52&$-$10.68&$-$10.52&$-$13.60&...&$-$14.03&32.50&29.96&29.85&29.28&29.78\\
1997&Mar&\phantom{0}5.1&\phantom{0}57.6&4.51&$-$10.33&...&$-$10.55&$-$10.67&$-$10.46&$-$14.10&$-$13.77&$-$13.70&32.32&...&29.85&29.21&29.77\\
1997&Mar&\phantom{0}5.1&\phantom{0}57.6&4.51&$-$10.39&$-$10.95&$-$10.54&$-$10.68&$-$10.49&$-$14.07&$-$13.77&$-$13.53&32.26&30.12&29.86&29.20&29.74\\
1997&Mar&\phantom{0}5.1&114.7&4.81&\phantom{0}$-$9.97&$-$10.70&$-$10.19&$-$10.64&$-$10.15&$-$13.65&$-$13.77&$-$13.44&32.69&30.37&30.20&29.24&30.08\\
1997&Jun&\phantom{0}4.2&146.7&5.02&...&...&$-$10.99&...&$-$11.00&...&...&$-$14.30&...&...&29.76&...&29.73\\
\vspace{0.05in}
1997&Jul&\phantom{0}1.2&114.7&4.98&...&...&$-$11.41&...&$-$11.46&$-$14.61&...&$-$14.36&...&...&29.61&...&29.53\\
2007&Dec&\phantom{0}4.2&\phantom{0}97.2&4.58&$-$10.38&$-$11.31&$-$10.77&$-$10.83&$-$10.68&{\it und}&$-$14.19&$-$14.53&31.93&29.58&29.45&28.92&29.42\\
2008&Mar&\phantom{0}4.2&\phantom{0}97.2&4.52&\phantom{0}$-$9.94&$-$10.79&$-$10.17&$-$10.48&$-$10.23&$-$13.77&$-$13.53&$-$13.69&32.13&29.76&29.66&29.03&29.62\\
2008&Mar&\phantom{0}4.2&\phantom{0}62.4&4.33&$-$10.22&$-$11.08&$-$10.44&$-$10.65&$-$10.51&$-$14.05&$-$13.82&$-$13.76&31.85&29.48&29.39&28.86&29.34\\
2008&Mar&\phantom{0}5.2&\phantom{0}77.8&4.43&$-$10.11&$-$10.97&$-$10.34&$-$10.59&$-$10.40&$-$13.99&$-$13.79&$-$13.77&31.96&29.59&29.50&28.92&29.46\\
\vspace{0.05in}
2008&Apr&30.2&\phantom{0}77.8&4.61&$-$10.94&$-$11.93&$-$11.25&$-$11.40&$-$11.33&$-$14.72&$-$14.49&$-$14.59&31.42&29.33&29.26&28.74&29.15\\
2018&Oct&\phantom{0}6.3&\phantom{0}97.2&4.19&$-$10.49&$-$11.44&$-$10.85&$-$10.65&$-$10.78&$-$14.57&$-$13.70&$-$13.80&31.10&28.71&28.62&28.36&28.57\\
2018&Oct&\phantom{0}6.3&204.5&4.51&$-$10.03&$-$10.90&$-$10.38&$-$10.31&$-$10.30&$-$14.04&$-$13.46&$-$13.50&31.56&29.25&29.09&28.69&29.05\\
2018&Oct&\phantom{0}6.3&\phantom{0}48.6&3.89&$-$10.96&$-$11.92&$-$11.31&$-$11.10&$-$11.27&$-$14.34&$-$13.92&$-$13.94&30.63&28.23&28.16&27.91&28.09\\
2018&Oct&\phantom{0}6.3&\phantom{0}97.2&4.19&$-$10.52&$-$11.46&$-$10.85&$-$10.68&$-$10.79&$-$14.20&$-$13.73&$-$13.75&31.07&28.69&28.62&28.33&28.57\\
2018&Nov&\phantom{0}7.2&204.5&4.25&\phantom{0}$-$9.34&$-$10.22&\phantom{0}$-$9.66&\phantom{0}$-$9.60&\phantom{0}$-$9.62&$-$13.16&$-$12.73&$-$12.80&31.66&29.25&29.10&28.74&29.07\\
2018&Nov&\phantom{0}7.3&\phantom{0}97.2&3.93&\phantom{0}$-$9.84&$-$10.73&$-$10.14&\phantom{0}$-$9.97&$-$10.10&$-$13.33&$-$13.00&$-$13.03&31.16&28.74&28.62&28.37&28.58\\
2018&Nov&\phantom{0}7.3&\phantom{0}48.6&3.63&$-$10.35&$-$11.23&$-$10.63&$-$10.33&$-$10.61&$-$13.76&$-$13.26&$-$13.28&30.65&28.23&28.13&28.00&28.08\\
2018&Nov&\phantom{0}7.3&\phantom{0}97.2&3.93&\phantom{0}$-$9.83&$-$10.72&$-$10.14&\phantom{0}$-$9.95&$-$10.10&$-$13.45&$-$13.00&$-$13.06&31.17&28.74&28.62&28.39&28.58\\
2018&Nov&15.3&155.9&4.05&\phantom{0}$-$9.48&$-$10.24&\phantom{0}$-$9.70&\phantom{0}$-$9.57&\phantom{0}$-$9.67&$-$13.22&$-$12.64&$-$12.67&31.44&29.01&28.86&28.57&28.82\\
2018&Nov&15.3&\phantom{0}48.6&3.54&$-$10.33&$-$11.11&$-$10.51&$-$10.22&$-$10.50&$-$13.61&$-$13.13&$-$13.15&30.59&28.15&28.05&27.92&27.99\\
2018&Nov&15.3&\phantom{0}97.2&3.84&\phantom{0}$-$9.83&$-$10.60&$-$10.02&\phantom{0}$-$9.81&$-$10.00&$-$13.33&$-$12.85&$-$12.89&31.09&28.65&28.53&28.33&28.49\\
2018&Nov&15.3&155.9&4.05&\phantom{0}$-$9.50&$-$10.26&\phantom{0}$-$9.70&\phantom{0}$-$9.58&\phantom{0}$-$9.68&$-$13.20&$-$12.70&$-$12.72&31.42&29.00&28.85&28.56&28.81\\
2018&Nov&15.3&\phantom{0}97.2&3.84&\phantom{0}$-$9.82&$-$10.60&$-$10.03&\phantom{0}$-$9.83&$-$10.01&$-$13.32&$-$12.86&$-$12.90&31.10&28.65&28.53&28.31&28.48\\
2018&Dec&\phantom{0}3.1&\phantom{0}97.2&3.59&\phantom{0}$-$9.60&$-$10.44&\phantom{0}$-$9.86&\phantom{0}$-$9.52&\phantom{0}$-$9.74&$-$12.83&$-$12.42&$-$12.46&30.76&28.34&28.24&28.07&28.20\\
2018&Dec&\phantom{0}3.1&204.5&3.92&\phantom{0}$-$9.10&\phantom{0}$-$9.89&\phantom{0}$-$9.35&\phantom{0}$-$9.11&\phantom{0}$-$9.21&$-$12.53&$-$12.12&$-$12.17&31.27&28.89&28.75&28.49&28.73\\
2018&Dec&\phantom{0}3.1&\phantom{0}48.6&3.29&$-$10.13&$-$10.97&$-$10.39&\phantom{0}$-$9.96&$-$10.27&$-$13.13&$-$12.71&$-$12.74&30.23&27.81&27.71&27.63&27.67\\
2018&Dec&\phantom{0}3.1&155.9&3.80&\phantom{0}$-$9.29&$-$10.08&\phantom{0}$-$9.53&\phantom{0}$-$9.25&\phantom{0}$-$9.40&$-$12.64&$-$12.23&$-$12.27&31.08&28.70&28.57&28.34&28.54\\
2018&Dec&\phantom{0}3.1&126.7&3.71&\phantom{0}$-$9.44&$-$10.24&\phantom{0}$-$9.69&\phantom{0}$-$9.37&\phantom{0}$-$9.55&$-$12.72&$-$12.30&$-$12.35&30.93&28.54&28.41&28.22&28.39\\
2018&Dec&\phantom{0}3.2&\phantom{0}77.8&3.50&\phantom{0}$-$9.78&$-$10.62&$-$10.03&\phantom{0}$-$9.65&\phantom{0}$-$9.91&$-$12.94&$-$12.51&$-$12.54&30.58&28.16&28.07&27.95&28.03\\
2018&Dec&\phantom{0}3.2&\phantom{0}62.4&3.40&\phantom{0}$-$9.97&$-$10.81&$-$10.21&\phantom{0}$-$9.81&$-$10.09&$-$13.02&$-$12.61&$-$12.64&30.39&27.97&27.89&27.78&27.85\\
2018&Dec&\phantom{0}3.2&\phantom{0}38.5&3.19&$-$10.33&$-$11.17&$-$10.55&$-$10.12&$-$10.45&$-$13.21&$-$12.81&$-$12.82&30.04&27.61&27.55&27.47&27.49\\
2018&Dec&\phantom{0}3.2&\phantom{0}24.5&2.99&$-$10.69&$-$11.55&$-$10.90&$-$10.45&$-$10.82&$-$13.42&$-$13.01&$-$13.03&29.68&27.23&27.20&27.15&27.12\\
2018&Dec&\phantom{0}3.2&\phantom{0}48.6&3.29&$-$10.16&$-$10.99&$-$10.37&\phantom{0}$-$9.95&$-$10.28&$-$13.13&$-$12.73&$-$12.73&30.20&27.79&27.73&27.65&27.66\\
2018&Dec&\phantom{0}3.2&\phantom{0}97.2&3.59&\phantom{0}$-$9.65&$-$10.44&\phantom{0}$-$9.88&\phantom{0}$-$9.52&\phantom{0}$-$9.76&$-$12.85&$-$12.42&$-$12.45&30.71&28.34&28.22&28.07&28.18\\
2018&Dec&\phantom{0}3.3&\phantom{0}97.2&3.59&\phantom{0}$-$9.63&$-$10.45&\phantom{0}$-$9.87&\phantom{0}$-$9.53&\phantom{0}$-$9.75&$-$12.87&$-$12.43&$-$12.45&30.73&28.33&28.23&28.07&28.19\\
2018&Dec&16.3&\phantom{0}97.2&3.43&\phantom{0}$-$9.65&$-$10.42&\phantom{0}$-$9.83&\phantom{0}$-$9.41&\phantom{0}$-$9.70&$-$12.39&$-$11.99&$-$11.99&30.42&28.02&27.99&27.86&27.92\\
2018&Dec&16.3&155.9&3.64&\phantom{0}$-$9.31&$-$10.05&\phantom{0}$-$9.49&\phantom{0}$-$9.13&\phantom{0}$-$9.35&$-$12.19&$-$11.80&$-$11.82&30.76&28.39&28.32&28.14&28.27\\
2018&Dec&16.3&\phantom{0}48.6&3.13&$-$10.17&$-$10.95&$-$10.38&\phantom{0}$-$9.90&$-$10.22&$-$12.70&$-$12.30&$-$12.31&29.90&27.49&27.44&27.37&27.40\\
2018&Dec&16.4&155.9&3.64&\phantom{0}$-$9.29&$-$10.03&\phantom{0}$-$9.49&\phantom{0}$-$9.13&\phantom{0}$-$9.34&$-$12.18&$-$11.79&$-$11.80&30.79&28.41&28.32&28.14&28.28\\
2018&Dec&30.1&\phantom{0}97.2&3.60&\phantom{0}$-$9.60&$-$10.46&\phantom{0}$-$9.77&\phantom{0}$-$9.63&\phantom{0}$-$9.84&$-$12.75&$-$12.36&$-$12.40&30.61&28.23&28.19&28.00&28.13\\
2018&Dec&30.1&204.5&3.92&\phantom{0}$-$9.08&\phantom{0}$-$9.91&\phantom{0}$-$9.24&\phantom{0}$-$9.21&\phantom{0}$-$9.30&$-$12.48&$-$12.09&$-$12.12&31.14&28.78&28.72&28.41&28.67\\
2018&Dec&30.1&\phantom{0}48.6&3.30&$-$10.11&$-$10.99&$-$10.27&$-$10.07&$-$10.35&$-$13.07&$-$12.66&$-$12.69&30.10&27.71&27.69&27.56&27.62\\
2018&Dec&30.1&\phantom{0}97.2&3.60&\phantom{0}$-$9.63&$-$10.48&\phantom{0}$-$9.77&\phantom{0}$-$9.63&\phantom{0}$-$9.84&$-$12.78&$-$12.39&$-$12.42&30.59&28.21&28.19&27.99&28.13\\
2019&Jan&31.2&\phantom{0}97.2&4.03&$-$10.04&$-$11.09&$-$10.35&$-$10.35&$-$10.42&$-$13.59&$-$13.21&$-$13.25&30.95&28.62&28.61&28.26&28.53\\
2019&Jan&31.2&204.5&4.36&\phantom{0}$-$9.54&$-$10.58&\phantom{0}$-$9.86&$-$10.01&\phantom{0}$-$9.91&$-$13.24&$-$12.93&$-$12.98&31.45&29.12&29.10&28.60&29.05\\
2019&Feb&26.2&\phantom{0}97.2&4.27&$-$10.51&$-$11.80&$-$10.91&$-$10.89&$-$10.99&$-$14.44&$-$13.78&$-$13.84&30.94&28.53&28.68&28.32&28.57\\
2019&Feb&26.3&\phantom{0}97.2&4.27&$-$10.51&$-$11.81&$-$10.91&$-$10.92&$-$11.00&$-$14.12&$-$13.85&$-$14.07&30.94&28.52&28.68&28.29&28.56\\
2019&Mar&25.2&\phantom{0}62.4&4.28&$-$11.44&$-$13.10&$-$11.68&$-$11.69&$-$11.77&$-$14.69&$-$14.65&$-$14.51&30.53&27.77&28.43&28.05&28.32\\
2019&Mar&25.2&\phantom{0}97.2&4.47&$-$11.07&$-$12.33&$-$11.39&$-$11.40&$-$11.48&$-$14.70&$-$14.42&$-$14.49&30.89&28.53&28.72&28.34&28.61\\
\enddata
\tablenotetext{a} {{\it ``und''} stands for ``undefined'' and means the continuum flux was measured but was less than 0.}
\label{t:phot_flux}
\end{deluxetable*}

\begin{deluxetable*}{lllccc@{\extracolsep{4pt}}ccccc@{\extracolsep{4pt}}ccc@{\extracolsep{4pt}}c}  
\tabletypesize{\scriptsize}
\tablecolumns{15}
\tablewidth{0pt} 
\setlength{\tabcolsep}{0.02in}
\tablecaption{Photometric production rates for comet 46P/Wirtanen.}
\tablehead{\multicolumn{3}{c}{UT Date}&
  \colhead{$\Delta$T}&
  \colhead{log $r_\mathrm{H}$}&
  \colhead{log $\rho$}&
  \multicolumn{5}{c}{log $Q$\tablenotemark{\scriptsize{a}}\phantom{00}(molecules s$^{-1}$)}&
  \multicolumn{3}{c}{log $A$($\theta$)$f\rho$\tablenotemark{\scriptsize{b}}\phantom{0}(cm)}&
  \colhead{log $Q$} \\
  \cline{7-11}
  \cline{12-14}
  \cline{15-15}
  \colhead{}&
  \colhead{}&
  \colhead{}&
  \colhead{(day)}&
  \colhead{(AU)}&
  \colhead{(km)}&
  \colhead{OH}&
  \colhead{NH}&
  \colhead{CN}&
  \colhead{C$_3$}&
  \colhead{C$_2$}&
  \colhead{UV}&
  \colhead{Blue}&
  \colhead{Green}&
  \colhead{H$_2$O}
  }
\startdata
\vspace{0.05in}
1991&Oct&11.5&\phantom{0}+21.0&\phantom{$-$}0.048&4.26&27.99{\tiny\phantom{.}.02}&25.86{\tiny\phantom{.}.04}&25.40{\tiny\phantom{.}.01}&24.78{\tiny\phantom{.}.02}&25.49{\tiny\phantom{.}.01}&1.88{\tiny\phantom{.}.08}&...{\tiny\phantom{.00}}&1.97{\tiny\phantom{.}.02}&28.10\\
1997&Feb&12.1&\phantom{0}$-$30.4&\phantom{$-$}0.056&4.63&28.07{\tiny\phantom{.}.10}&25.56{\tiny\phantom{.}.15}&25.18{\tiny\phantom{.}.02}&24.71{\tiny\phantom{.}.09}&25.27{\tiny\phantom{.}.02}&1.67{\tiny\phantom{.}.36}&...{\tiny\phantom{.00}}&1.54{\tiny\phantom{.}.15}&28.18\\
1997&Feb&15.1&\phantom{0}$-$27.4&\phantom{$-$}0.051&4.63&28.00{\tiny\phantom{.}.04}&25.62{\tiny\phantom{.}.09}&25.22{\tiny\phantom{.}.01}&24.86{\tiny\phantom{.}.04}&25.35{\tiny\phantom{.}.01}&2.18{\tiny\phantom{.}.10}&...{\tiny\phantom{.00}}&1.52{\tiny\phantom{.}.06}&28.10\\
1997&Mar&\phantom{0}5.1&\phantom{0}\phantom{0}$-$9.4&\phantom{$-$}0.030&4.51&27.96{\tiny\phantom{.}.03}&...{\tiny\phantom{.00}}&25.36{\tiny\phantom{.}.01}&24.87{\tiny\phantom{.}.03}&25.47{\tiny\phantom{.}.01}&1.82{\tiny\phantom{.}.15}&1.82{\tiny\phantom{.}.07}&1.91{\tiny\phantom{.}.05}&28.08\\
1997&Mar&\phantom{0}5.1&\phantom{0}\phantom{0}$-$9.4&\phantom{$-$}0.030&4.51&27.90{\tiny\phantom{.}.04}&25.94{\tiny\phantom{.}.04}&25.37{\tiny\phantom{.}.01}&24.86{\tiny\phantom{.}.03}&25.44{\tiny\phantom{.}.01}&1.76{\tiny\phantom{.}.12}&0.00{\tiny\phantom{.}.07}&2.07{\tiny\phantom{.}.03}&28.02\\
1997&Mar&\phantom{0}5.1&\phantom{0}\phantom{0}$-$9.4&\phantom{$-$}0.030&4.81&27.97{\tiny\phantom{.}.05}&25.79{\tiny\phantom{.}.06}&25.39{\tiny\phantom{.}.01}&24.79{\tiny\phantom{.}.06}&25.48{\tiny\phantom{.}.01}&1.88{\tiny\phantom{.}.15}&0.00{\tiny\phantom{.}.07}&1.86{\tiny\phantom{.}.04}&28.09\\
1997&Jun&\phantom{0}4.2&\phantom{0}+81.7&\phantom{$-$}0.174&5.02&...{\tiny\phantom{.00}}&...{\tiny\phantom{.00}}&24.74{\tiny\phantom{.}.03}&...{\tiny\phantom{.00}}&24.91{\tiny\phantom{.}.03}&...{\tiny\phantom{.00}}&...{\tiny\phantom{.00}}&1.28{\tiny\phantom{.}.19}&...\\
\vspace{0.05in}
1997&Jul&\phantom{0}1.2&+108.7&\phantom{$-$}0.236&4.98&...{\tiny\phantom{.00}}&...{\tiny\phantom{.00}}&24.64{\tiny\phantom{.}.05}&...{\tiny\phantom{.00}}&24.75{\tiny\phantom{.}.08}&1.51{\tiny\phantom{.}.20}&...{\tiny\phantom{.00}}&1.52{\tiny\phantom{.}.20}&...\\
2007&Dec&\phantom{0}4.2&\phantom{0}$-$60.3&\phantom{$-$}0.121&4.58&27.53{\tiny\phantom{.}.03}&25.38{\tiny\phantom{.}.10}&24.92{\tiny\phantom{.}.12}&24.46{\tiny\phantom{.}.11}&25.08{\tiny\phantom{.}.01}&{\it und}&1.20{\tiny\phantom{.}.11}&0.88{\tiny\phantom{.}.10}&27.60\\
2008&Mar&\phantom{0}4.2&\phantom{0}+30.7&\phantom{$-$}0.055&4.52&27.76{\tiny\phantom{.}.00}&25.58{\tiny\phantom{.}.01}&25.16{\tiny\phantom{.}.00}&24.66{\tiny\phantom{.}.01}&25.32{\tiny\phantom{.}.00}&1.76{\tiny\phantom{.}.05}&1.67{\tiny\phantom{.}.03}&1.54{\tiny\phantom{.}.03}&27.87\\
2008&Mar&\phantom{0}4.2&\phantom{0}+30.7&\phantom{$-$}0.055&4.33&27.76{\tiny\phantom{.}.01}&25.59{\tiny\phantom{.}.01}&25.14{\tiny\phantom{.}.00}&24.61{\tiny\phantom{.}.01}&25.29{\tiny\phantom{.}.00}&1.67{\tiny\phantom{.}.06}&1.57{\tiny\phantom{.}.04}&1.65{\tiny\phantom{.}.03}&27.86\\
2008&Mar&\phantom{0}5.2&\phantom{0}+31.7&\phantom{$-$}0.057&4.43&27.72{\tiny\phantom{.}.01}&25.55{\tiny\phantom{.}.01}&25.12{\tiny\phantom{.}.00}&24.60{\tiny\phantom{.}.01}&25.27{\tiny\phantom{.}.00}&1.64{\tiny\phantom{.}.06}&1.52{\tiny\phantom{.}.05}&1.56{\tiny\phantom{.}.03}&27.83\\
\vspace{0.05in}
2008&Apr&30.2&\phantom{0}+87.8&\phantom{$-$}0.188&4.61&27.04{\tiny\phantom{.}.06}&25.15{\tiny\phantom{.}.10}&24.72{\tiny\phantom{.}.03}&24.22{\tiny\phantom{.}.09}&24.81{\tiny\phantom{.}.03}&1.36{\tiny\phantom{.}.40}&1.26{\tiny\phantom{.}.20}&1.18{\tiny\phantom{.}.18}&27.08\\
2018&Oct&\phantom{0}6.3&\phantom{0}$-$67.6&\phantom{$-$}0.139&4.19&27.31{\tiny\phantom{.}.03}&25.15{\tiny\phantom{.}.04}&24.65{\tiny\phantom{.}.05}&24.21{\tiny\phantom{.}.04}&24.80{\tiny\phantom{.}.01}&0.80{\tiny\phantom{.}.38}&1.34{\tiny\phantom{.}.05}&1.26{\tiny\phantom{.}.05}&27.37\\
2018&Oct&\phantom{0}6.3&\phantom{0}$-$67.6&\phantom{$-$}0.139&4.51&27.27{\tiny\phantom{.}.02}&25.17{\tiny\phantom{.}.02}&24.66{\tiny\phantom{.}.02}&24.27{\tiny\phantom{.}.02}&24.81{\tiny\phantom{.}.01}&1.00{\tiny\phantom{.}.25}&1.25{\tiny\phantom{.}.05}&1.24{\tiny\phantom{.}.05}&27.33\\
2018&Oct&\phantom{0}6.3&\phantom{0}$-$67.6&\phantom{$-$}0.139&3.89&27.33{\tiny\phantom{.}.04}&25.18{\tiny\phantom{.}.05}&24.66{\tiny\phantom{.}.02}&24.11{\tiny\phantom{.}.03}&24.79{\tiny\phantom{.}.01}&1.33{\tiny\phantom{.}.13}&1.41{\tiny\phantom{.}.04}&1.42{\tiny\phantom{.}.04}&27.39\\
2018&Oct&\phantom{0}6.3&\phantom{0}$-$67.6&\phantom{$-$}0.139&4.19&27.27{\tiny\phantom{.}.02}&25.13{\tiny\phantom{.}.04}&24.65{\tiny\phantom{.}.02}&24.19{\tiny\phantom{.}.02}&24.79{\tiny\phantom{.}.01}&1.17{\tiny\phantom{.}.18}&1.31{\tiny\phantom{.}.04}&1.30{\tiny\phantom{.}.04}&27.34\\
2018&Nov&\phantom{0}7.2&\phantom{0}$-$35.7&\phantom{$-$}0.064&4.25&27.69{\tiny\phantom{.}.01}&25.49{\tiny\phantom{.}.01}&24.97{\tiny\phantom{.}.00}&24.55{\tiny\phantom{.}.01}&25.13{\tiny\phantom{.}.00}&1.47{\tiny\phantom{.}.05}&1.57{\tiny\phantom{.}.01}&1.53{\tiny\phantom{.}.01}&27.79\\
2018&Nov&\phantom{0}7.3&\phantom{0}$-$35.7&\phantom{$-$}0.064&3.93&27.69{\tiny\phantom{.}.01}&25.51{\tiny\phantom{.}.01}&24.97{\tiny\phantom{.}.00}&24.50{\tiny\phantom{.}.01}&25.13{\tiny\phantom{.}.00}&1.62{\tiny\phantom{.}.04}&1.62{\tiny\phantom{.}.01}&1.62{\tiny\phantom{.}.01}&27.79\\
2018&Nov&\phantom{0}7.3&\phantom{0}$-$35.7&\phantom{$-$}0.064&3.63&27.68{\tiny\phantom{.}.02}&25.52{\tiny\phantom{.}.02}&24.97{\tiny\phantom{.}.01}&24.50{\tiny\phantom{.}.01}&25.12{\tiny\phantom{.}.00}&1.49{\tiny\phantom{.}.05}&1.67{\tiny\phantom{.}.01}&1.66{\tiny\phantom{.}.01}&27.78\\
2018&Nov&\phantom{0}7.3&\phantom{0}$-$35.6&\phantom{$-$}0.064&3.93&27.70{\tiny\phantom{.}.01}&25.52{\tiny\phantom{.}.01}&24.97{\tiny\phantom{.}.01}&24.51{\tiny\phantom{.}.01}&25.13{\tiny\phantom{.}.00}&1.51{\tiny\phantom{.}.05}&1.63{\tiny\phantom{.}.01}&1.59{\tiny\phantom{.}.01}&27.81\\
2018&Nov&15.3&\phantom{0}$-$27.7&\phantom{$-$}0.049&4.05&27.76{\tiny\phantom{.}.01}&25.56{\tiny\phantom{.}.01}&25.01{\tiny\phantom{.}.00}&24.57{\tiny\phantom{.}.00}&25.16{\tiny\phantom{.}.00}&1.41{\tiny\phantom{.}.04}&1.67{\tiny\phantom{.}.01}&1.66{\tiny\phantom{.}.01}&27.87\\
2018&Nov&15.3&\phantom{0}$-$27.7&\phantom{$-$}0.049&3.54&27.74{\tiny\phantom{.}.02}&25.56{\tiny\phantom{.}.01}&25.00{\tiny\phantom{.}.01}&24.52{\tiny\phantom{.}.01}&25.15{\tiny\phantom{.}.00}&1.53{\tiny\phantom{.}.03}&1.69{\tiny\phantom{.}.01}&1.69{\tiny\phantom{.}.01}&27.85\\
2018&Nov&15.3&\phantom{0}$-$27.6&\phantom{$-$}0.049&3.84&27.74{\tiny\phantom{.}.01}&25.55{\tiny\phantom{.}.01}&25.00{\tiny\phantom{.}.00}&24.54{\tiny\phantom{.}.00}&25.15{\tiny\phantom{.}.00}&1.51{\tiny\phantom{.}.03}&1.66{\tiny\phantom{.}.01}&1.65{\tiny\phantom{.}.01}&27.85\\
2018&Nov&15.3&\phantom{0}$-$27.6&\phantom{$-$}0.049&4.05&27.74{\tiny\phantom{.}.01}&25.55{\tiny\phantom{.}.01}&25.00{\tiny\phantom{.}.00}&24.56{\tiny\phantom{.}.00}&25.15{\tiny\phantom{.}.00}&1.44{\tiny\phantom{.}.04}&1.61{\tiny\phantom{.}.01}&1.61{\tiny\phantom{.}.01}&27.85\\
2018&Nov&15.3&\phantom{0}$-$27.6&\phantom{$-$}0.049&3.84&27.75{\tiny\phantom{.}.02}&25.55{\tiny\phantom{.}.01}&24.99{\tiny\phantom{.}.00}&24.53{\tiny\phantom{.}.01}&25.15{\tiny\phantom{.}.00}&1.53{\tiny\phantom{.}.04}&1.66{\tiny\phantom{.}.01}&1.63{\tiny\phantom{.}.01}&27.86\\
2018&Dec&\phantom{0}3.1&\phantom{0}\phantom{0}$-$9.8&\phantom{$-$}0.027&3.59&27.80{\tiny\phantom{.}.01}&25.64{\tiny\phantom{.}.01}&25.08{\tiny\phantom{.}.00}&24.59{\tiny\phantom{.}.00}&25.25{\tiny\phantom{.}.00}&1.72{\tiny\phantom{.}.01}&1.80{\tiny\phantom{.}.00}&1.78{\tiny\phantom{.}.00}&27.92\\
2018&Dec&\phantom{0}3.1&\phantom{0}\phantom{0}$-$9.8&\phantom{$-$}0.027&3.92&27.78{\tiny\phantom{.}.00}&25.64{\tiny\phantom{.}.00}&25.08{\tiny\phantom{.}.00}&24.62{\tiny\phantom{.}.00}&25.26{\tiny\phantom{.}.00}&1.69{\tiny\phantom{.}.01}&1.77{\tiny\phantom{.}.00}&1.75{\tiny\phantom{.}.00}&27.90\\
2018&Dec&\phantom{0}3.1&\phantom{0}\phantom{0}$-$9.8&\phantom{$-$}0.027&3.29&27.79{\tiny\phantom{.}.01}&25.64{\tiny\phantom{.}.01}&25.06{\tiny\phantom{.}.00}&24.58{\tiny\phantom{.}.00}&25.23{\tiny\phantom{.}.00}&1.71{\tiny\phantom{.}.01}&1.80{\tiny\phantom{.}.01}&1.80{\tiny\phantom{.}.01}&27.91\\
2018&Dec&\phantom{0}3.1&\phantom{0}\phantom{0}$-$9.8&\phantom{$-$}0.027&3.80&27.78{\tiny\phantom{.}.00}&25.64{\tiny\phantom{.}.00}&25.08{\tiny\phantom{.}.00}&24.61{\tiny\phantom{.}.00}&25.25{\tiny\phantom{.}.00}&1.70{\tiny\phantom{.}.01}&1.79{\tiny\phantom{.}.00}&1.76{\tiny\phantom{.}.00}&27.90\\
2018&Dec&\phantom{0}3.1&\phantom{0}\phantom{0}$-$9.8&\phantom{$-$}0.027&3.71&27.78{\tiny\phantom{.}.00}&25.64{\tiny\phantom{.}.00}&25.07{\tiny\phantom{.}.00}&24.59{\tiny\phantom{.}.00}&25.25{\tiny\phantom{.}.00}&1.71{\tiny\phantom{.}.01}&1.80{\tiny\phantom{.}.00}&1.77{\tiny\phantom{.}.00}&27.90\\
2018&Dec&\phantom{0}3.2&\phantom{0}\phantom{0}$-$9.8&\phantom{$-$}0.027&3.50&27.79{\tiny\phantom{.}.00}&25.63{\tiny\phantom{.}.01}&25.08{\tiny\phantom{.}.00}&24.60{\tiny\phantom{.}.00}&25.24{\tiny\phantom{.}.00}&1.71{\tiny\phantom{.}.01}&1.80{\tiny\phantom{.}.00}&1.79{\tiny\phantom{.}.00}&27.91\\
2018&Dec&\phantom{0}3.2&\phantom{0}\phantom{0}$-$9.8&\phantom{$-$}0.027&3.40&27.76{\tiny\phantom{.}.01}&25.60{\tiny\phantom{.}.01}&25.05{\tiny\phantom{.}.00}&24.57{\tiny\phantom{.}.00}&25.22{\tiny\phantom{.}.00}&1.72{\tiny\phantom{.}.01}&1.80{\tiny\phantom{.}.00}&1.79{\tiny\phantom{.}.00}&27.88\\
2018&Dec&\phantom{0}3.2&\phantom{0}\phantom{0}$-$9.8&\phantom{$-$}0.027&3.19&27.78{\tiny\phantom{.}.01}&25.62{\tiny\phantom{.}.01}&25.07{\tiny\phantom{.}.00}&24.57{\tiny\phantom{.}.00}&25.22{\tiny\phantom{.}.00}&1.73{\tiny\phantom{.}.01}&1.81{\tiny\phantom{.}.01}&1.82{\tiny\phantom{.}.01}&27.90\\
2018&Dec&\phantom{0}3.2&\phantom{0}\phantom{0}$-$9.8&\phantom{$-$}0.027&2.99&27.76{\tiny\phantom{.}.01}&25.59{\tiny\phantom{.}.01}&25.06{\tiny\phantom{.}.01}&24.56{\tiny\phantom{.}.01}&25.20{\tiny\phantom{.}.01}&1.73{\tiny\phantom{.}.01}&1.81{\tiny\phantom{.}.01}&1.81{\tiny\phantom{.}.01}&27.88\\
2018&Dec&\phantom{0}3.2&\phantom{0}\phantom{0}$-$9.7&\phantom{$-$}0.027&3.29&27.76{\tiny\phantom{.}.01}&25.61{\tiny\phantom{.}.01}&25.08{\tiny\phantom{.}.00}&24.59{\tiny\phantom{.}.00}&25.22{\tiny\phantom{.}.00}&1.71{\tiny\phantom{.}.01}&1.79{\tiny\phantom{.}.00}&1.81{\tiny\phantom{.}.01}&27.88\\
2018&Dec&\phantom{0}3.2&\phantom{0}\phantom{0}$-$9.7&\phantom{$-$}0.027&3.59&27.75{\tiny\phantom{.}.00}&25.64{\tiny\phantom{.}.00}&25.07{\tiny\phantom{.}.00}&24.59{\tiny\phantom{.}.00}&25.23{\tiny\phantom{.}.00}&1.69{\tiny\phantom{.}.01}&1.80{\tiny\phantom{.}.00}&1.79{\tiny\phantom{.}.00}&27.87\\
2018&Dec&\phantom{0}3.3&\phantom{0}\phantom{0}$-$9.7&\phantom{$-$}0.027&3.59&27.77{\tiny\phantom{.}.00}&25.63{\tiny\phantom{.}.00}&25.07{\tiny\phantom{.}.00}&24.59{\tiny\phantom{.}.00}&25.23{\tiny\phantom{.}.00}&1.67{\tiny\phantom{.}.01}&1.79{\tiny\phantom{.}.00}&1.79{\tiny\phantom{.}.00}&27.89\\
2018&Dec&16.3&\phantom{0}\phantom{0}+3.3&\phantom{$-$}0.024&3.43&27.73{\tiny\phantom{.}.00}&25.59{\tiny\phantom{.}.00}&25.09{\tiny\phantom{.}.00}&24.60{\tiny\phantom{.}.00}&25.23{\tiny\phantom{.}.00}&1.99{\tiny\phantom{.}.00}&2.06{\tiny\phantom{.}.00}&2.08{\tiny\phantom{.}.00}&27.85\\
2018&Dec&16.3&\phantom{0}\phantom{0}+3.4&\phantom{$-$}0.024&3.64&27.72{\tiny\phantom{.}.00}&25.61{\tiny\phantom{.}.00}&25.08{\tiny\phantom{.}.00}&24.60{\tiny\phantom{.}.00}&25.24{\tiny\phantom{.}.00}&1.98{\tiny\phantom{.}.00}&2.05{\tiny\phantom{.}.00}&2.05{\tiny\phantom{.}.00}&27.84\\
2018&Dec&16.3&\phantom{0}\phantom{0}+3.4&\phantom{$-$}0.024&3.13&27.74{\tiny\phantom{.}.00}&25.59{\tiny\phantom{.}.01}&25.05{\tiny\phantom{.}.00}&24.56{\tiny\phantom{.}.00}&25.22{\tiny\phantom{.}.00}&1.98{\tiny\phantom{.}.00}&2.05{\tiny\phantom{.}.00}&2.07{\tiny\phantom{.}.00}&27.86\\
2018&Dec&16.4&\phantom{0}\phantom{0}+3.4&\phantom{$-$}0.024&3.64&27.75{\tiny\phantom{.}.00}&25.62{\tiny\phantom{.}.00}&25.08{\tiny\phantom{.}.00}&24.59{\tiny\phantom{.}.00}&25.24{\tiny\phantom{.}.00}&1.99{\tiny\phantom{.}.00}&2.06{\tiny\phantom{.}.00}&2.07{\tiny\phantom{.}.00}&27.87\\
2018&Dec&30.1&\phantom{0}+17.1&\phantom{$-$}0.034&3.60&27.65{\tiny\phantom{.}.01}&25.52{\tiny\phantom{.}.01}&25.03{\tiny\phantom{.}.00}&24.51{\tiny\phantom{.}.00}&25.17{\tiny\phantom{.}.00}&1.81{\tiny\phantom{.}.01}&1.88{\tiny\phantom{.}.00}&1.86{\tiny\phantom{.}.00}&27.77\\
2018&Dec&30.1&\phantom{0}+17.2&\phantom{$-$}0.034&3.92&27.64{\tiny\phantom{.}.00}&25.53{\tiny\phantom{.}.00}&25.04{\tiny\phantom{.}.00}&24.53{\tiny\phantom{.}.00}&25.19{\tiny\phantom{.}.00}&1.76{\tiny\phantom{.}.01}&1.83{\tiny\phantom{.}.00}&1.82{\tiny\phantom{.}.00}&27.76\\
2018&Dec&30.1&\phantom{0}+17.2&\phantom{$-$}0.034&3.30&27.66{\tiny\phantom{.}.01}&25.53{\tiny\phantom{.}.01}&25.04{\tiny\phantom{.}.00}&24.50{\tiny\phantom{.}.00}&25.17{\tiny\phantom{.}.00}&1.80{\tiny\phantom{.}.01}&1.88{\tiny\phantom{.}.00}&1.87{\tiny\phantom{.}.01}&27.78\\
2018&Dec&30.1&\phantom{0}+17.2&\phantom{$-$}0.034&3.60&27.63{\tiny\phantom{.}.00}&25.50{\tiny\phantom{.}.00}&25.02{\tiny\phantom{.}.00}&24.51{\tiny\phantom{.}.00}&25.17{\tiny\phantom{.}.00}&1.78{\tiny\phantom{.}.01}&1.85{\tiny\phantom{.}.00}&1.84{\tiny\phantom{.}.00}&27.74\\
2019&Jan&31.2&\phantom{0}+49.2&\phantom{$-$}0.094&4.03&27.35{\tiny\phantom{.}.00}&25.25{\tiny\phantom{.}.01}&24.83{\tiny\phantom{.}.00}&24.28{\tiny\phantom{.}.01}&24.95{\tiny\phantom{.}.00}&1.53{\tiny\phantom{.}.02}&1.59{\tiny\phantom{.}.01}&1.57{\tiny\phantom{.}.01}&27.44\\
2019&Jan&31.2&\phantom{0}+49.3&\phantom{$-$}0.094&4.36&27.35{\tiny\phantom{.}.00}&25.23{\tiny\phantom{.}.01}&24.84{\tiny\phantom{.}.00}&24.31{\tiny\phantom{.}.01}&24.98{\tiny\phantom{.}.00}&1.56{\tiny\phantom{.}.02}&1.54{\tiny\phantom{.}.01}&1.51{\tiny\phantom{.}.01}&27.43\\
2019&Feb&26.2&\phantom{0}+75.2&\phantom{$-$}0.157&4.27&27.04{\tiny\phantom{.}.01}&24.85{\tiny\phantom{.}.04}&24.60{\tiny\phantom{.}.01}&24.09{\tiny\phantom{.}.02}&24.69{\tiny\phantom{.}.01}&1.05{\tiny\phantom{.}.15}&1.38{\tiny\phantom{.}.04}&1.34{\tiny\phantom{.}.04}&27.10\\
2019&Feb&26.3&\phantom{0}+75.3&\phantom{$-$}0.157&4.27&27.04{\tiny\phantom{.}.01}&24.85{\tiny\phantom{.}.03}&24.60{\tiny\phantom{.}.01}&24.06{\tiny\phantom{.}.02}&24.68{\tiny\phantom{.}.01}&1.37{\tiny\phantom{.}.07}&1.31{\tiny\phantom{.}.04}&1.11{\tiny\phantom{.}.07}&27.09\\
2019&Mar&25.2&+102.2&\phantom{$-$}0.221&4.28&26.69{\tiny\phantom{.}.02}&24.16{\tiny\phantom{.}.24}&24.41{\tiny\phantom{.}.02}&23.82{\tiny\phantom{.}.06}&24.49{\tiny\phantom{.}.04}&1.31{\tiny\phantom{.}.15}&1.02{\tiny\phantom{.}.16}&1.19{\tiny\phantom{.}.11}&26.71\\
2019&Mar&25.2&+102.3&\phantom{$-$}0.221&4.47&26.75{\tiny\phantom{.}.01}&24.61{\tiny\phantom{.}.10}&24.41{\tiny\phantom{.}.02}&23.91{\tiny\phantom{.}.05}&24.49{\tiny\phantom{.}.03}&1.11{\tiny\phantom{.}.23}&1.07{\tiny\phantom{.}.14}&1.02{\tiny\phantom{.}.16}&26.77\\
\enddata
\tablenotetext{a} {Production rates and \afrhot\ followed by the upper, i.e. the positive, uncertainty. The ``+'' and ``$-$'' uncertainties are equal as percentages, but unequal in log-space; the ``$-$'' values can be computed.}
\tablenotetext{b} {{\it ``und''} stands for ``undefined'' and means the continuum flux was measured but was less than 0.}
\label{t:phot_rates}
\end{deluxetable*}

\renewcommand{\baselinestretch}{1.0} 

\vspace{-23mm}
\subsection{Gas Production Rates vs Distance, Time, and Apparition}
It is immediately evident from Figure~\ref{fig:photometry} that the overall quality of the data from 
the newest apparition is much better than that from prior years, and permits us to 
readily examine the slopes of the production rates with heliocentric distance. 
In fact the {\rh}-dependencies are unusually linear in log-log 
space. In particular, the carbon-bearing species (CN, C$_3$, and C$_2$) exhibit very similar 
slopes (ranging from $-$3.5 to $-$3.7 to $-$3.9, respectively) and very little difference 
before and after perihelion. In comparison, both OH and NH vary in a similar manner -- 
while only slightly steeper inbound ($-$4.3 and $-$4.2, respectively) than the carbon-bearing 
species, they fall-off significantly faster outbound ($-$5.1 and $-$5.2, if one ignores the 
abnormally low NH point on the final night [off the plot in Fig.~\ref{fig:photometry}]). These results strongly imply a compositional 
heterogeneity between the two source regions identified in the modeling from Section~\ref{sec:modeling}. 
Several other comets have exhibited differing behavior in the carbon-bearing 
species and the hydrogen-bearing species, including 2P/Encke \citep{ahearn83b,ahearn95}
and 45P/Honda-Mrkos-Pajusakova (our database). 
However, we emphasize that the composition of both regions on the surface of 
Wirtanen fall within the ``typical'' class of our large database \citep{schleicher16b}; 
the ratio of carbon-bearing species to water can vary by more than a factor of two 
and still be in the typical class. 

It is also evident from examination of Figure~\ref{fig:photometry} 
that gas production rates were significantly lower in the most recent apparition. 
However, before examining this decrease in more detail, we first looked for possible 
aperture trends that might have an effect on our derived production rates, because 
Wirtanen approached Earth so much closer than ever before. Looking first at the night of 
December 3 with its nine aperture sizes, ranging from 24 to 204 arcsec ($\rho$ = 980 to 8320 km), 
and then other nights near closest approach, we see no trends with aperture size for 
OH, NH, or CN. Even C$_3$ and C$_2$ exhibit small but consistent trends (an increase 
of only 15\% for a factor of 8$\times$ in aperture radius). Larger apertures early and 
late in the apparition exhibit no trends at all. Therefore, we conclude that the 
lower gas production rates measured in 2018/19 are real. 

Allowing for the observed heliocentric distance trends, when they could be reliably 
determined, we see an ongoing secular decrease across all four apparitions. For instance, 
CN decreases by $\sim$28\% from 1991 to 1997, $\sim$21\% from 1997 to 2008, and by $\sim$31\% 
from 2008 to 2018, for a total decrease of about 60\%. C$_2$ and C$_3$ are similar, decreasing by totals 
of about 53\% and 46\%, respectively. Due to the pre-/post-perihelion asymmetries previously 
noted for OH and NH, the secular trends are somewhat suppressed, but yet are still $\sim$51\% 
and $\sim$53\%, respectively. This factor of two or more decrease is the second largest 
that we've measured within our entire photometry database of more than 190 comets 
\citep{schleicher16b}, exceeded only by 103P/Hartley 2 -- another hyperactive object 
having only recently had its perihelion distance substantially decreased. 

\subsection{Dust Behavior}
The nominal {\afrhot} values for the green continuum shown in the lower-left 
panel of Figure~\ref{fig:photometry} exhibit much more apparent scatter than do the molecular species 
because of two reasons -- phase angle effects and significant aperture trends -- 
that must be corrected in order to investigate the underlying behavior of the dust grains. 
As is often the case for the dust in comets, there is a clear downward trend in the 
derived {\afrhot} values with increasing aperture size, which has in the past been 
attributed to some type of fading grains, likely either due to grains becoming 
darker or shrinking in size with time as they coast away from the nucleus \citep[cf.][]{baum92}. 

\begin{figure}
  \centering
  \includegraphics[width=85mm]{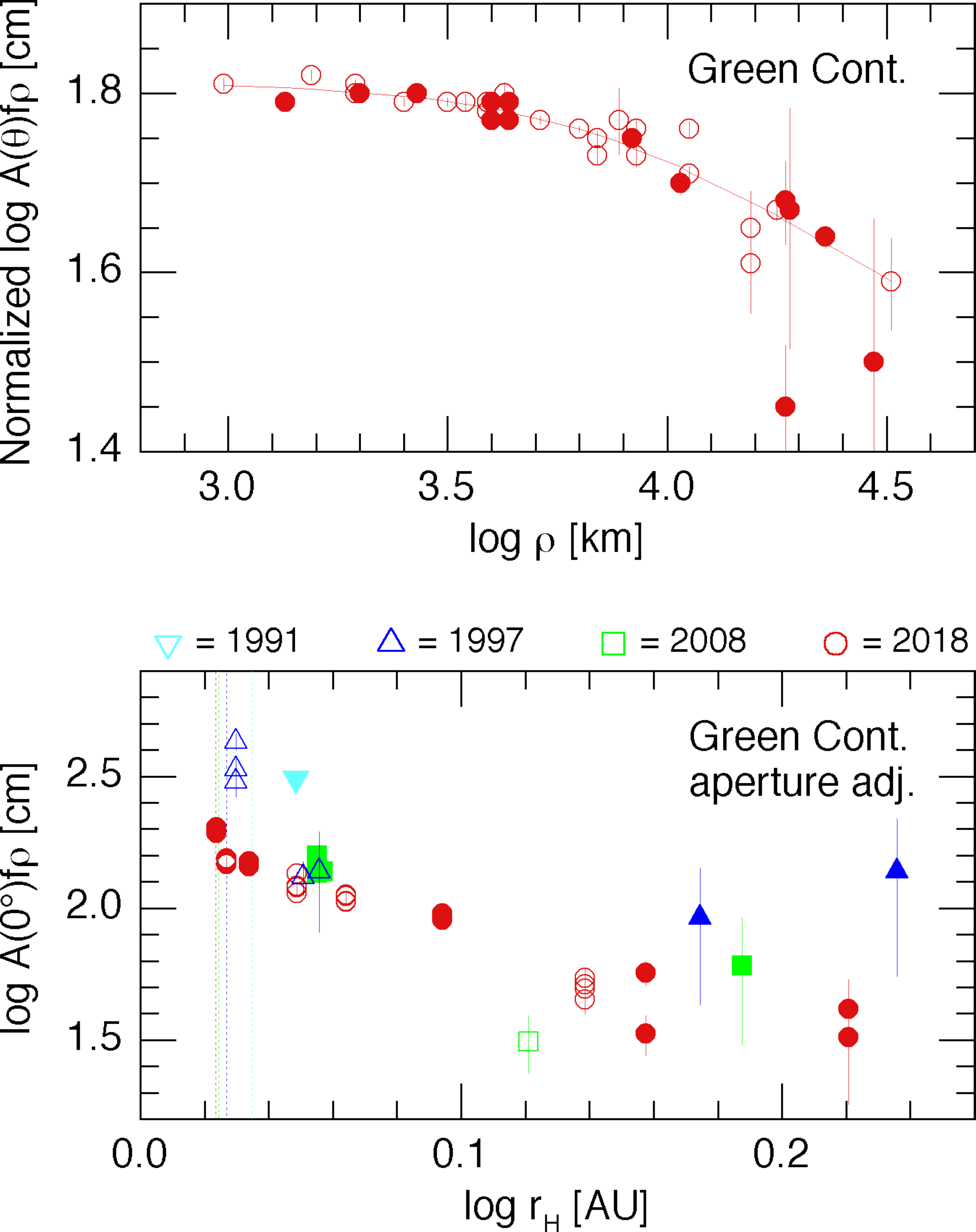}
  \caption{{\it Top:} Logarithm of normalized $A(\theta)f\rho$ for the dust as a function of the logarithm of the aperture radius for the 2018/19 apparition. {\it Bottom:} $A(0^{\circ})f\rho$ for the dust as a function of the logarithm of the heliocentric distance. Symbols for both panels are as given in Figure~\ref{fig:photometry}. The curve in the top panel is the composite aperture adjustment described in the text.}
  \label{fig:afrho}
\end{figure}

Accounting for this aperture effect is particularly important due to the very wide range 
in aperture sizes employed during the 2018/19 apparition associated with the very 
close approach of Wirtanen to the Earth. Fortunately, the characteristics of this 
aperture effect were consistent during the apparition. We, therefore, used a multi-step 
procedure to adjust all {\afrhot} results to a normalized aperture radius of $10^4$~km. 
Because our best sampling took place on December 3, with 12 observational sets over nine different 
apertures, we used that night's results of log {\afrhot} vs log $\rho$ as our fiducial. 
We then adjusted the data from each other night up or down in log space to best overlay the December 3 data, 
ultimately yielding the compound aperture curve shown in the top panel of Figure~\ref{fig:afrho}. 
This curve was then re-normalized at $10^4$~km, and each original {\afrhot} result 
was adjusted appropriately. Because the measurements at earlier apparitions generally 
were made with much larger projected aperture sizes due to Wirtanen's much larger 
geocentric distances, we also applied a first-order aperture correction to these 
{\afrhot} results, simply using a near-linear (in log space) extrapolation of the  
aperture correction curve just discussed, again normalizing to $10^4$~km. 

We next applied our standard correction for phase angle effects \citep[cf.][]{schleicher11}, 
normalizing to 0$^\circ$ phase angle 
(the phase adjustment for each night is given along with the phase angle in Table~\ref{t:phot_circ}). 
The results, after applying both adjustments, are shown in the bottom panel of 
Figure~\ref{fig:afrho}. The {\rh}-dependence ($-$3.8) for 2018/19 is most similar to 
that of C$_2$, discussed previously. It is clear that there has also been a major 
decrease in dust production, by more than a factor of two, from the 1990s to 
the past dozen years, though possibly not at the same rate of change from orbit 
to orbit as for the gas species. This last conclusion is uncertain, however, due to 
its dependence on the aperture extrapolation for large aperture sizes noted above. 

The color of the dust grains, based on the green and UV continuum data, is 
unexceptional, with a mean value of $\Delta$log({\afrhoz}) of $\sim$0.09, while the green minus blue
value is about $-$0.01; each is somewhat below usual. We also detect a very small trend 
in color with aperture size. Note however, that the low dust-to-gas ratio 
in Wirtanen, discussed next, implies the possibility of residual contamination 
of the continuum bandpasses by various gas emissions, even following our standard 
decontamination procedures. Our experience with the entire photometry database suggests 
that both the more neutral colors and the slight aperture trend in colors could readily 
be due to contaminations, and we therefore do not pursue colors further. 

The mean log dust-to-gas ratio, based on log[{\afrhoz}] $-$ log $Q$(OH), is $-$25.55 
cm s molecule$^{-1}$, only about one-quarter of the phase adjusted value in comets 
1P/Halley, 81P/Wild 2, and 67P/Churyumov-Gerasimenko, or one-half the value in 9P/Tempel 1, i.e., somewhat 
less dusty than comets explored by spacecraft but not at all unusual; 
for instance it is 15$\times$ greater than the extreme case of 2P/Encke \citep{schleicher16b}.

\subsection{Water Production and Effective Active Area}
\label{sec:water_prod}

Using our empirical conversion from a Haser OH production rate to the equivalent 
vectorial water production rate \citep[cf.][]{cochran93,schleicher98}, 
we obtain the values listed in the final column in Table~\ref{t:phot_rates}. With a factor 
of $r_\mathrm{H}^{0.5}$ in the conversion, the final slopes are 0.5 steeper than the 
values listed for OH above. Given the observed asymmetry about perihelion, it is not 
surprising that the peak value for water, $8{\times}10^{27}$ molecules s$^{-1}$, was measured on 
December 3 (${\Delta}T= -10$~day) rather than on our next night of photometry of December 16. 
Using a water vaporization model based on the work of \citet[][with a factor of two 
error corrected, M.\ A'Hearn 2010 private communication]{cowan79}, this peak 
value corresponds to an effective active area of about 2.5~km$^2$. Combined with an 
effective nucleus radius of 0.6~km (\citealt{lamy98}; consistent with radar estimates, E.~Howell, priv. comm.), one obtains an active fractional area of 
about 55\%, as compared to a corrected value of about 80\% in 1997 \citep{farnham98}. 
There are several caveats, however, regarding this value. 
First, the area and active fraction are significantly smaller than this at larger 
heliocentric distances because of the steep {\rh}-dependence; at the furthest point 
out-bound, the values drop to only one-fifth of the peak values. Second, the imaging 
results strongly suggest that some, if not the majority, of the water is coming off of the nucleus 
as icy grains (see next section). 
In this case, the source regions could be significantly smaller than 
areas computed with the vaporization model. Finally, the peak values in the 1990s 
slightly exceed the nominal total surface area of 4.5 km$^2$, further suggesting an 
icy grain component for the observed water.

A comparison of water production rate determinations by various investigators for 
the 1997 apparition, when Wirtanen was the planned {\it Rosetta} target, has 
already been performed by \citet{fink04}.
As they discussed in detail, there is considerable scatter among 
the various datasets but most of this apparent disagreement is a result of  
investigators using differing values for Haser scalelengths. 
The most extensive dataset, both during 1997 and most subsequent apparitions, 
was obtained using the SWAN instrument on {\it SOHO} as recently summarized 
by \cite{combi21}. In addition to the only measurements during the 2002 apparition, 
they also obtained data in 2008 and 2018. Besides using very different techniques, 
with SWAN measuring the very large hydrogen Lyman-$\alpha$ coma while we measure the inner OH 
coma, the cadence of observations greatly differs. While SWAN's brightness 
threshold permits only a small range of heliocentric distances to be covered, 
it can often obtain near-daily measurements when the comet is within range. 
Thus, our {\rh}-dependent slopes are more robust while SWAN can be used to identify 
outbursts. Keeping these aspects in mind, we refer to Figure 2 in \cite{combi21} 
where there is a very large dispersion among their measurements during the 2002 
apparition. In fact, only five of the 28 data points might be considered baseline 
activity and very close to the 1997 values, while the remaining are as much as 
$5{\times}-6\times$ higher and appear to be associated with several sustained outbursts. 
In 2008 we have no overlap in datasets, as ours are all at larger distances than 
the SWAN data, though an extrapolation of our values is in reasonable agreement. 

Of the five nights near perihelion for which we have data close in time with the 
SWAN results, we start somewhat higher in mid-November, are quite close in early-December, 
about 10\%\ low in mid- and late-December, and quite close at the end of January, though 
comparisons are inexact due to night-to-night variations in the SWAN data (\citealt{combi21}, 
see their Figure 2). 
We conclude that our agreement is quite good overall, despite sometimes large 
differences in derived {\rh}-dependencies between our values and theirs 
that we attribute to our generally having much wider ranges of heliocentric distance. 
In particular, we note that they found that water production in Wirtanen dropped 
by about a factor of two between 1997 and 2018, essentially identical to our own 
findings.

\section{IMAGING RESULTS}
\label{sec:imaging}

In contrast to our just discussed photometry which spanned multiple apparitions, our imaging covered only the 2018/19 apparition, but with many more epochs of observations in order to glean as much information as possible out of the rapidly varying viewing geometry. 
As is evident from Table~\ref{t:imaging_circ}, the majority of our imaging nights consisted of only CN and broadband {\it R} or {\it r}$'$ (to monitor dust) images. This was the result of several factors. OH, NH, and C$_3$ are fainter than CN, 
and OH and NH are subject to significant atmospheric extinction that limits their observability to only very low airmasses. Furthermore, NH, C$_2$, and C$_3$ suffer dust contamination that required photometric conditions in order to be properly analyzed, and such calibrations were too time consuming except on our dedicated nights. We made a programmatic decision to prioritize regular CN monitoring from December onwards so that we could tightly constrain the rotation period at as many epochs as possible to look for a changing rotation period; this strategy has proven to be very successful in the past, e.g., for C/2004 Q2 Machholz \citep{farnham07a}, C/2007 N3 Lulin \citep{bair18}, 103P/Hartley 2 \citep{knight11b,samarasinha11}, and 41P/Tuttle-Giacobini-Kres\'ak \citep{schleicher19}. 
The CN imaging was the focus of Paper I; below we concentrate on the morphology and behavior of the dust and the remaining gas species.

\subsection{Dust Morphology and Behavior}
\label{sec:dust_morph}

Although our imaging emphasis throughout the apparition was on the gas species, we obtained frequent broadband {\it R} or {\it r}$'$ images since they have minimal gas contamination, making them a high signal-to-noise proxy for a pure dust image. Narrowband continuum filters were used on many photometric nights and confirmed the validity of this approach, so we will henceforth refer to {\it R} or {\it r}$'$ images as ``dust'' images. We nearly always paired dust images with CN images, meaning that we have a robust dataset for searching for rotational variation in the dust as well as the gas coma. 

Throughout the apparition, dust images were brighter in the anti-sunward hemisphere, with the brightness concentrated in the expected tailward direction (see the top row of Figure~\ref{fig:gas_morph}). 
The general shape evolved slowly as the viewing geometry changed, but did not change appreciably on shorter, e.g., rotational, timescales. We did not detect any fainter, time-varying features analogous to the gas jets (discussed in Paper I and the following subsection), despite analyzing the dust images with the same enhancement techniques. 
However, we note that we generally ignore features within a few PSFs of the center in enhanced images since this region is most susceptible to processing artifacts; any real and varying dust features, if they exist, likely extend less than $\sim$5 arcsec, or $\lesssim$1000 km. The lone exception occurred on December 16 and 17, when a persistent feature appeared very close to the nucleus. We attribute this to either slow moving dust from the December 12 outburst \citep{farnham21} or to a quirk of the viewing geometry; see Section~\ref{sec:overall_model} for further discussion.

The lack of an obvious rotational signature in the dust is both expected and surprising. On the one hand, we have detected time-varying jet-like features in CN in many comets brighter than $V{\sim}13$ and within $\sim$1 AU of Earth over the years with only a few, e.g., C/1996 B2 Hyakutake \citep{schleicher03a} and 103P/Hartley 2 \citep{knight13a}, exhibiting rotational variation in the dust, so the lack of rotation signature in Wirtanen is not surprising. On the other hand, we have generally assumed that such features are not seen, in part, because the dust has low ejection velocities and large velocity dispersions, requiring excellent spatial resolution to be identified. 
We have also frequently seen approximately stationary sunward-facing features in some comets, e.g., C/2007 N3 Lulin \citep{bair18}, 41P/Tuttle-Giacobini-Kres\'ak \citep{bair18,schleicher19}, yet such features are absent in Wirtanen. 
Wirtanen's historically close apparition should have made the identification of any such features far easier than in most comets.  

There is a faint, but noticeable, asymmetry in the bulk dust brightness which favors the hemisphere in which the carbon-bearing gases (CN, C$_2$, and C$_3$) are brightest. This is most obvious in Figure~\ref{fig:gas_morph} on December 13 and 16, where these gases are brightest to the southeast, and the dust images are clearly brighter to the southeast than to the northwest. We attribute this asymmetry to the majority of the dust originating from the same source regions as the gases before being pushed tailward by radiation pressure. We will further explore the ties between the dust and gas in Section~\ref{sec:discussion}.

\subsection{Gas Morphology}
\label{sec:gas_morph}

\begin{figure*}
  \centering
    \includegraphics[width=180mm]{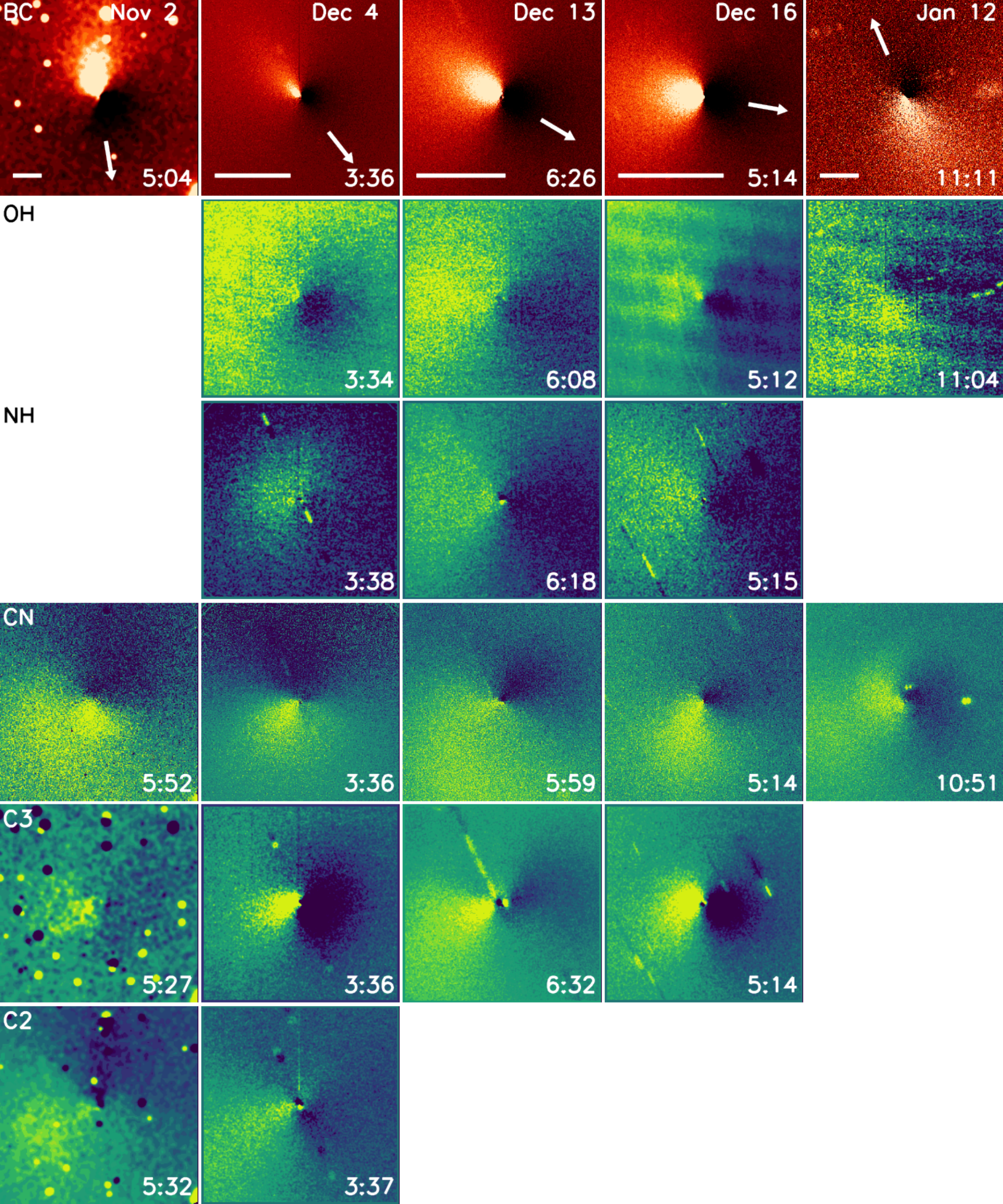}
  \caption{{\footnotesize Morphology of dust and pure gas species at five different epochs. The species shown on a given row is labeled in the first column, and the date of a given column is displayed in the top panel. Image mid-times are in the bottom right corner.
ll images are centered on the nucleus with north up and east left, and have been enhanced by division of an azimuthal median. A 5000 km scale bar and the direction to the Sun are indicated on the top row for each night. Trailed stars can be seen in many images; positive (yellow) are stars in the gas image while negative (blue) are stars in the continuum image that was removed. Faint vertical lines are seen in some panels due to a bad column on the LMI detector, while the regular, cross-hatched pattern in the OH panels is due to the construction of the CCD which becomes visible in faint UV images.}
}
  \label{fig:gas_morph}
\end{figure*}

We obtained images of sufficient quality to assess the gas morphology in November (CN, C$_3$, C$_2$), December (OH, NH, CN, C$_2$, and C$_3$), January (OH, CN), and February (CN). Representative pure gas images following continuum subtraction as well as a BC (dust) image on five nights are shown in Figure~\ref{fig:gas_morph}. These images have all been enhanced by subtraction of an azimuthal median (cf.\ \citealt{schleicher04,samarasinha14}); the CN morphology differs somewhat from images at the same times shown in Paper I because in that paper we removed a temporal average to enhance the images. Temporal averaging requires evenly spaced images in rotational phase of similar quality and with minimal intervening change in viewing geometry. As there were insufficient images of any other gas species to construct the needed temporal average, we have opted to show all images here with the same enhancement technique for direct comparison. As discussed in Paper~I, the temporal average enhancement made faint features in the darker half of the image more evident; without this enhancement here we cannot comment on the existence of such features in OH, NH, C$_2$, or C$_3$.

An additional difference in appearance relative to Paper I is our removal of underlying dust continuum when possible. The amount of underlying continuum varies from species to species (cf.\ \citealt{farnham00}), but is increasingly less important at larger distances from the nucleus since dust brightness falls off much more rapidly than does gas brightness. As described in the previous subsection, throughout most of the apparition the only distinct dust feature was the tail. This provided a convenient way to check that the continuum removal worked as intended, since over- or under-removal would result in negative or positive tail signatures remaining in the gas images, respectively. 
The only potential concern is the broad OH and NH morphology which mostly overlaps the continuum. However, in both cases the gas lacks an obvious increase in brightness overlapping the tail. As the tail is much brighter than the bulk brightness at comparable distances from the nucleus, its signature should be evident in the ``pure'' gas images if our calibrations had gone awry. We can be confident in our assessments of the gas jets described below since these do not appear in the continuum and are unlikely to be an artifact of the calibration process.

Since CN had the highest signal-to-noise on all nights, we use it as our fiducial for comparison to the other species. The CN morphology gradually evolved during the apparition as described in detail in Paper~I and shown in their Figures~3 and 6. We briefly summarize it here to provide the context for the other species that did not have comparable temporal coverage. There were two obvious CN jets seen during the apparition, one that persisted throughout a rotation and one 
whose brightness was more strongly modulated. The morphology progressed from a face-on spiral rotating in the clockwise direction in November to a broad, edge-on spiral with hints of a corkscrew structure and little variation with rotational phase for most of the rotation cycle in mid-December, to a face-on spiral with the reverse sense of rotation (i.e., counter-clockwise) by January. 
The morphology evolution will be further revisited in the next section as a modeling constraint.

As also discussed in Paper~I and shown in their Figure~6, the CN morphology repeated with a period of $\sim$9 hr, with the period first increasing then decreasing during the apparition. For the following comparisons we have chosen the CN image whose midpoint is closest in time to the other species being considered. Since the gas images presented here were generally median combinations of 3--5 images, and some nights' images were acquired sequentially while others were interspersed, the mid-times are sometimes within a minute of a CN frame, while on other nights, they differ by up to 45 minutes. Even in the most extreme case, the temporal separation is relatively small compared to the period of repetition and does not affect our interpretation. Furthermore, given the high-fidelity of our model (Section~\ref{sec:modeling}) to replicate the CN morphology throughout the entire apparition, we can confidently predict what the CN morphology would have been at the time of another image as an additional check.

\subsubsection{Carbon-bearing species: CN, C$_2$, and C$_3$}
CN, C$_2$, and C$_3$ exhibit similar morphology when acquired contemporaneously, as can be seen in Figure~\ref{fig:gas_morph}. On 2018 November 2 (first column) no jet is discerned, but all three are brightest in the southeast quadrant. On December 4, 13, and 16 (middle three columns), all three species have a dominant feature leaving the nucleus at a position angle (P.A.) measured from north through east around $120^{\circ}-135^\circ$. Neither C$_2$ nor C$_3$ was obtained after 2018 December, so they are not seen in the final column.  Although we only show one set per night, multiple sets were acquired on some nights and the C$_2$ and C$_3$ morphology varied in a similar manner to the CN from set to set. 
We thus conclude that all three carbon-bearing species originate from the same source regions, a conclusion supported by their similar production rate {\rh}-dependencies and explored further in Section~\ref{sec:discussion}.

Although the general appearance and locations of features are similar across CN, C$_2$, and C$_3$, they are not identical for several reasons. None of the three species is considered to be a ``parent'' volatile; they do not sublimate directly from the ice on the nucleus, but instead arise as a second generation (``daughter'') or third generation (``grand-daughter'') produced via photodissociation or chemical processes. The parent(s) or grandparent(s) of the three species are not well constrained, but CN and C$_2$ are likely primarily daughter species from several parents, while C$_3$ (and C$_2$ to some extent) may be produced by chemical reactions in the coma \citep[cf.][]{helbert05}. The combination of different formation pathways with different lifetimes -- C$_3$ has a Haser scale length roughly an order of magnitude shorter than CN and 2.4$\times$ shorter than C$_2$ \citep[cf.][]{ahearn95} -- results in different spatial distributions for each species. These natural differences in morphology are further muddled by the non-simultaneity of some images and, as will be shown in Section~\ref{sec:modeling}, the fact that the primary feature we see varying in December is actually composed of two wide, overlapping jets.

\subsubsection{OH and NH -- Products of icy grains?}
OH and NH exhibit similar morphology to each other, although the NH data are relatively limited, and both are different from that of the carbon-bearing species. We were not able to make an assessment in 2018 November as the signal-to-noise of OH was too low and NH was not acquired. As shown in the second through fifth columns of Figure~\ref{fig:gas_morph}, OH and NH both exhibited a roughly hemispheric brightness enhancement in the direction of the dust tail (shown in the BC row) and lacked any well-defined jets as seen in CN, C$_2$, and C$_3$. There was no evidence of variation with rotational phase for either OH or NH on any of these nights.

\begin{figure}
  \centering
    \includegraphics[width=84mm]{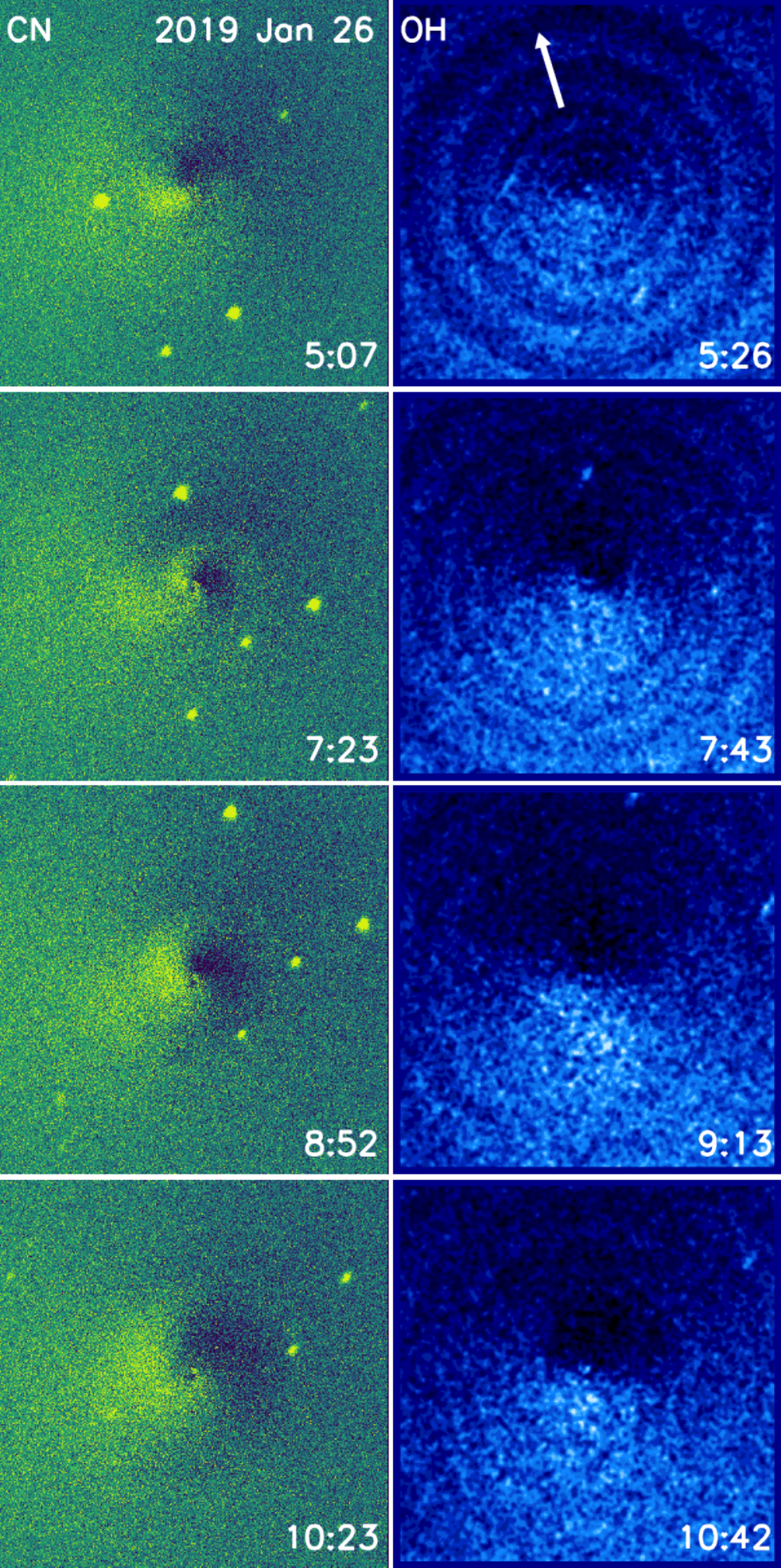}
  \caption{Comparison of CN and OH throughout one night (2019 Jan 26). CN images are plotted in the left column, OH in the right column. Image mid-times are given on each panel. All images are 60,000 km across at the comet, are centered on the nucleus, have been enhanced by subtraction of an azimuthal median, and are aligned with north up and east to the left. 
  Trailed stars can be seen in some CN frames, while the ring-like structures seen in some OH panels are artifacts of the enhancement.}
  \label{fig:cnoh_jan26}
\end{figure}

The rotational variation of CN was discussed thoroughly in Paper I, but we revisit it briefly here to further illustrate the invariance of OH. 
We obtained CN images with a $\sim$45 minute cadence for 9--11 hours on 2019 January 26 and 28 as well as OH images every $\sim$45 min for $\sim$5 hours on the same nights. The temporal coverage and spacing ensure a full rotational cycle for CN was sampled each night; when both nights are phased together we also have a complete OH rotational cycle, and repetition of CN morphology from night to night confirms the validity of this combination. 
As shown in Figure~\ref{fig:cnoh_jan26}, CN evinces a pinwheel shape, with one or two jets visible and rotating counter-clockwise throughout the night, and the bulk brightness is always primarily to the east. The OH shows no variation, but is always brightest towards the south on this night, in the general direction of the tail.

The OH and NH morphology and lack of rotational variation are consistent with these gases being released from icy grains in the coma. Icy grains containing OH and NH (or more likely their parent molecules) released from the nucleus would be swept tailward due to radiation pressure, and the excess velocities in random directions imparted during photodissociation, coupled with long lifetimes, would result in a much broader distribution of the resulting OH and NH than exhibited by the dust tail. Icy grains have long been hypothesized as existing in the comae of some comets \citep[e.g.,][]{ahearn84}. Their unequivocal detection by {\it EPOXI} in 103P/Hartley 2 \citep{ahearn11a,protopapa14} explained that comet's hyperactivity, so their existence in Wirtanen, also considered hyperactive, was anticipated before the apparition. Surprisingly, icy grains were not detected in Wirtanen via IR spectroscopy \citep{protopapa20}. 
We observed similar dimorphism between OH \& NH and the carbon-bearing species in Hartley 2 \citep{knight13a} as reported here for Wirtanen, strengthening our confidence in the icy grain explanation. 
Although atmospheric extinction makes obtaining high signal-to-noise OH observations challenging for most comets, the utility of OH imaging for identifying icy grains is clear. 

We hypothesize that the icy grains are released from the same source regions as the carbon-bearing species for several reasons. First, the OH and NH, like the dust, is somewhat brighter in the direction from which CN, C$_2$, and C$_3$ originate, as would be expected. The icy grains are presumably slow moving with large velocity dispersions, so any rotational signature is washed out, but the bulk brightness enhancement is preserved. Second, as will be discussed next, our modeling finds that the two source regions for the CN jets subtend $\sim$30\% of the surface, more than half of the 55\% active fraction we find from photometry. If the remaining water (having an equivalent active fraction of $\sim$25\% of the nucleus) was released from other source regions, we'd expect some sort of morphological signature, but none is seen. 
Lastly, as discussed previously the production rates of all five gas species behaved generally similarly, suggesting they all originated from the same source regions. However, the steeper post-perihelion decrease in OH and NH than the carbon bearing species does suggest some compositional heterogeneity, to which we will return in Section~\ref{sec:discussion}.

\section{MODELING THE CN JET MORPHOLOGY}
\label{sec:modeling}

\subsection{Modeling Overview}

Our Monte Carlo jet model has seen numerous enhancements since its early success 
in reproducing the dust jets observed in Comet Hyakutake \citep[C/1996 B2;][]{schleicher03a}. 
Most relevant here, we added the ability to vectorially apply the 
excess velocity of dissociation of a parent species to an observed daughter gas 
species, with the most recent example in our study of Comet Lulin \citep[C/2007 N3;][]{bair18}. 
Since most aspects of the model are described in detail in 
these papers, we only briefly summarize other key aspects of the model. The model 
allows us to control the orientation and rotation rate of the nucleus, the size 
and location of multiple source regions, and their illumination function, as a 
function of time throughout an apparition. While all computations are performed 
in the comet's reference frame, a series of coordinate transformations yield 
the view as seen by the observer on Earth.

In practice, we incorporate a systematic top-down search of the 
multi-dimensional parameter space by first varying the parameters that 
determine the basic shape of an observed jet -- pole orientation and source latitude -- 
for representative images throughout the apparition. 
Viable solutions are further constrained by examining full rotational sequences. 
Because the apparent shape of spiral jets can change substantially as the width of 
the jet varies from narrow to broad, the entire process is repeated using several 
different widths. Only after converging on the comet's pole orientation and each 
source location and size, are variations to other parameters investigated, 
such as the solar illumination function, outflow velocities, and the degree of 
dispersion from the radial direction. Finally, the values of all parameters are 
refined in an iterative process.

\subsection{Basic Constraints for Comet Wirtanen}

\begin{figure}
  \centering
    \includegraphics[width=85mm]{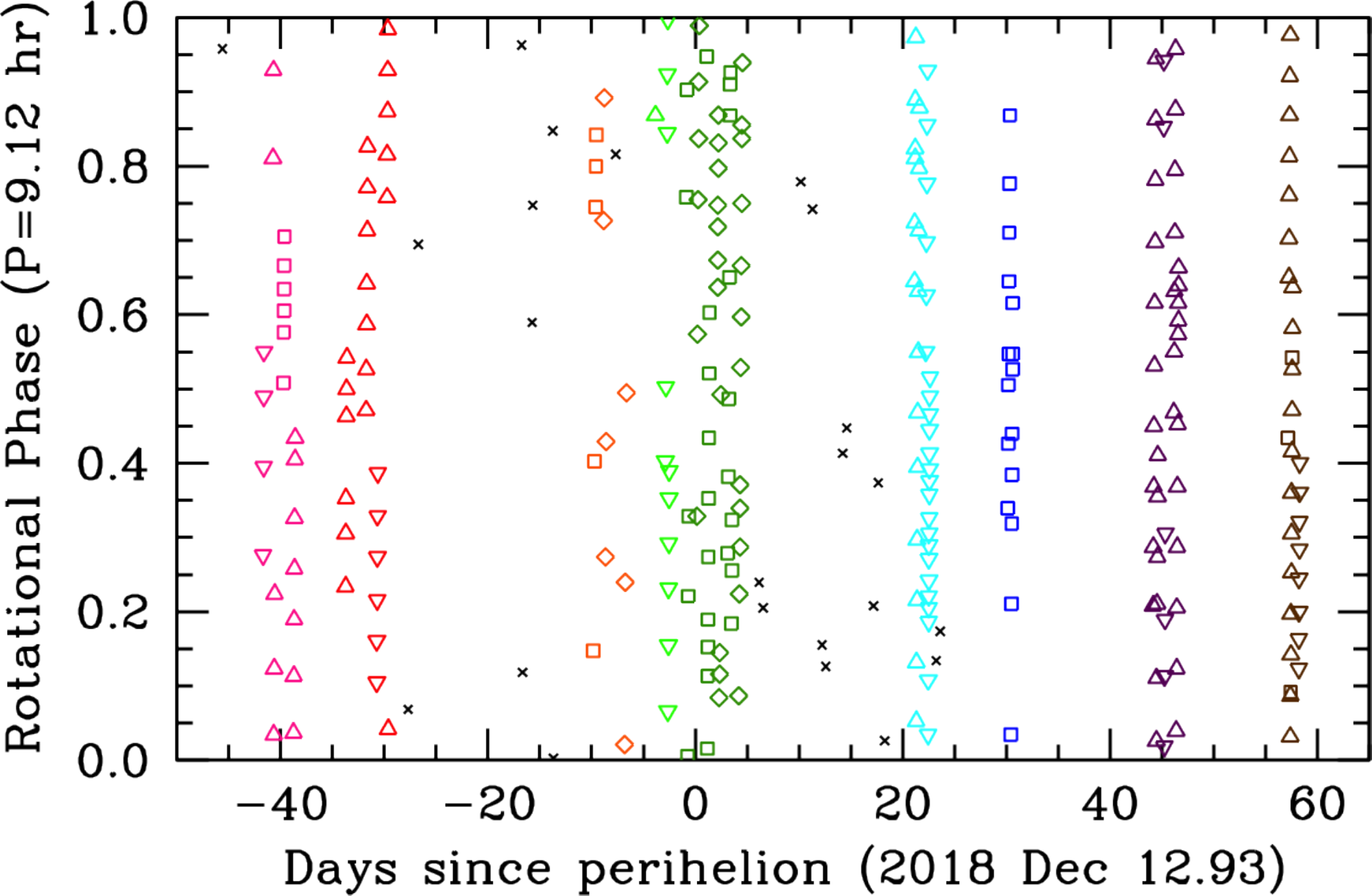}
  \caption{Rotational phase coverage of our CN images as a function of time 
from perihelion (${\Delta}T$) assuming a fixed sidereal rotation period of 9.12 hr. 
Squares and diamonds (on alternating nights to distinguish each night's data) are 
observations obtained with the LDT, while triangles (up and down) are for 42-inch 
data. Each color represents an observing run, while black crosses indicate snapshot 
images from various telescopes (see Table~\ref{t:imaging_circ}). Because Wirtanen's actual sidereal 
period varied significantly throughout the apparition, and only peaked at 9.14 hr 
in mid-December, this plot shows the distribution of observations in rotational 
phase space but not the actual rotational phases, which shift from run to run.}
  \label{fig:cn_phase_plot}
\end{figure}

As noted in the Introduction, Comet Wirtanen's 2018/19 apparition provided a
nearly ideal set of circumstances for detailed monitoring of any morphological 
features in the inner coma over a wide range of viewing geometries. However, 
as will be seen, some of Wirtanen's specific characteristics also provided 
several challenges during the process of modeling the observed jet morphology. 
We first itemize basic characteristics regarding Wirtanen's physical properties 
that directly impact the subsequent modeling, some of which were already described and 
discussed. First, little to no variation was observed 
for the dust, and therefore only the CN gas images are used in the model. 
Fortunately, the typical high contrast of CN emission to the underlying continuum 
combined with Wirtanen's relatively low dust-to-gas ratio, meant that there was 
no need to remove the continuum even quite close to the nucleus. Successive CN 
spiral features have similar brightnesses, shapes, and spacing, but subtle 
differences in an alternating pattern, thereby strongly suggesting two similar-sized 
source regions on opposite sides of the nucleus. The morphology is essentially 
identical from cycle to cycle and over multiple days, directly implying that 
there is no significant non-principal-axis rotation involved. Thus the $\sim$9 hr 
periodicity requires that the sidereal period is also $\sim$9 hr. 
To emphasize the extent of our dataset during this apparition, we plot the 
rotational phase coverage, as a function of time from perihelion for each set of 
CN images in Figure~\ref{fig:cn_phase_plot}.

The CN features exhibit definitive 
clockwise rotation in the first-half of November, and counter-clockwise rotation throughout 
January and early February; therefore, we must be viewing from opposite hemispheres
at these times and the changeover occurred near closest approach. The shape of 
these features along with coarse modeling directly implies that the obliquity 
of the rotation axis was near 90$^\circ$ and that both source regions were near the 
equator. However, specific differences in the shapes of successive jets, 
particularly on the anti-sunward side both early and late in the apparition, 
with one jet showing face-on spirals and the other jet transitioning between 
face-on and side-on (i.e. corkscrew) spirals, implies the sources are centered 
at somewhat different latitudes. The features are always relatively round in 
shape and appear shell-like even as our view crossed the equator near 
closest approach during the jets' transition from an apparent clockwise to 
counter-clockwise rotation, thereby requiring very broad jets. This characteristic 
also explains why a specific time for our passing through the plane of the jet 
is difficult to determine and why our line of sight seems to remain ``within'' the jet (in fact, both jets) 
for many days. Finally, the scale and outward motion of features imply that the bulk 
outflow velocity of the gas is $\sim$0.8 km s$^{-1}$.

\subsection{Additional Issues and Complications}

As described in Paper I, both the overall and specific appearance 
of the CN morphology can change greatly depending on which enhancement methodology 
is used. With a strong sunward/anti-sunward asymmetry usually present, 
features on the anti-sunward side are often difficult to detect with the 
azimuthal average/median removal, but are readily evident using the temporal 
average removal since bulk features present throughout a rotational cycle are 
eliminated. While our model cannot directly replicate this latter enhancement 
method, it can crudely match the result by ignoring the solar illumination and 
allowing all particles to be emitted independent of the location of the Sun, thereby 
effectively removing  the bulk asymmetry. Therefore, during our modeling we 
alternated between assuming a standard solar illumination function such as the 
cosine of the Sun's zenith distance when attempting to match the actual brightnesses 
and this alternative when fitting features on the ``night'' side. Examples shown 
later in this section give the azimuthal average removal and temporal removal of 
representative images, along with the model solution using the solar function. 
As modeling progressed, it became obvious that indeed some gas is released even 
when the Sun is below the local horizon, but at a much reduced rate (roughly 
10\%\ of the rate when the Sun is at the zenith); this is included in the final model. 

As noted, the shape of the features near perihelion need both jets to be very 
broad. Some aspects directly require that the source regions stretch over a large 
range of latitudes since the solar illumination function varies with latitude, 
while other aspects can be due to either a dispersion from the normal in the 
outflow direction and/or the amount of excess velocity in random directions 
when the CN parent dissociates. The latter in turn depends on the gas density 
and the collision rate -- in the collisional zone, daughter molecules will be forced 
to continue to move radially outward, mimicking the Haser model, while beyond 
the collision zone the daughters will behave as expected from the vectorial 
model. Here, we assume a small amount of initial dispersion from the local normal 
direction (10$^\circ$ radius Gaussian) but, due to Wirtanen's relatively low water production 
rate as compared to other comets that we've modeled, a higher than usual net 
excess daughter velocity (0.25 km s$^{-1}$). Note that these specific values are intertwined 
with the size of the source regions, and although this combination overall provides 
a good match of the entire coma brightness throughout the apparition, a moderate 
change in the value of one of these parameters can often be compensated for by an 
associated change in another of these parameters.  

While Wirtanen's broad and generally ``soft'' features increased the difficulty 
in performing our modeling, especially when attempting to match these features 
during the rapidly changing viewing geometry in December, another, more serious 
complication was caused by Wirtanen's changing rotation period. 
Our modeling confirmed the preliminary conclusion found in Paper I that the 
increase and subsequent decrease in Wirtanen's apparent period was too large to 
be explained by synodic effects alone. Simply based on the pole constraints noted 
in the previous sub-section, the maximum offset between apparent and synodic 
periods could be no more than $\sim$0.05 hr, whereas the apparent period changed 
by $\sim$0.20 hr between mid-December and early February. A major consequence 
is that no single sidereal period could reproduce the morphology without causing 
large changes in source longitudes early and late in the apparition. 
As a practical aspect, since all calculations have the time of 
perihelion as the origin for temporal and longitudinal calculations, not knowing 
{\it a priori} the functional form of the changes in sidereal period meant that 
offsets accumulated over time. Ultimately, in an iterative process, we determined 
average sidereal periods between perihelion and each observing run necessary to 
prevent computed source longitudes from drifting over time. We return to the 
results from this process later in this section.

\subsection{Constraints for Refining the Model Solution and Associated Results}
\label{sec:model_constraints}

During the overall iterative process of determining a final model solution, 
narrowing the range of viable pole orientations was a high priority. However, 
because of the large breadth of each jet, when we viewed the jets from a line of sight within, 
particles emitted from the extremes of the source region appeared to behave 
differently from the bulk of the particles. Thus the following is based both on 
the bulk characteristics of each jet as well as examining the shape and 
behavior of the extrema. As noted previously, in early November the jets clearly 
had an overall clockwise rotation, while in January the motion was counter-clockwise. 
Close examination revealed evidence of clockwise motion of the near-equatorial jet 
as late as during the December 3--6 run, and the opposite rotation by December 16--17. 
Combined, these suggest that the Earth crossed the equator, or the specific latitude 
of the center of the source region, about December 10--12. This, in turn, yields the 
principal angle of the pole (i.e., the perpendicular direction to the obliquity) of 
about 235--240$^\circ$, and is nearly independent of the obliquity. Separately, the motion of 
features with rotation in early to mid-December was approximately north-to-south, 
with a P.A. of about 0--20$^\circ$. These P.A.s near perihelion require 
that the obliquity was $<$75$^\circ$. By late January, the shapes of each jet place a lower 
limit on the obliquity of $\sim$60$^\circ$; thus, the obliquity of Wirtanen's nucleus 
was most likely in the range of 65--70$^\circ$. When combined with our final jet locations, 
our preferred solution, to the nearest 5$^\circ$, had an obliquity of 70$^\circ$ and a principal 
angle of 240$^\circ$ in Wirtanen's frame of reference. This directly corresponds to 
an R.A. of 319$^\circ$ and Declination of $-$5$^\circ$ and, due to our use of the righthand rule 
in the model, this is the direction of the positive or ``north'' pole.

To provide a crude check on the orientation of the comet's spin axis, we
turned to a basic technique that relies on measurements of coma features that
reflect the apparent pole orientation (e.g., persistent linear jets or
conical features centered around the spin axis).  We have successfully used
this technique to derive the pole orientations of a number of comets,
including spacecraft targets where in situ data confirmed our results
\citep[e.g.,][]{farnham02,farnham05,schleicher06b}. Using the
P.A. of the projected pole, we define a plane that contains
the P.A. vector and the line-of-sight vector; the spin axis
lies somewhere within this plane.  Applying this process to different
observations produces multiple planes, and if the viewing geometry changes
sufficiently between the observations, then the intersection of the various
planes reveals the orientation of the pole in three dimensions.

In the case of Wirtanen, we use the oscillating corkscrew shape observed to
the east in early- and mid-December \citep{farnham21}.  This
feature is characteristic of a jet emanating from a source at mid-latitudes,
with the radially outflowing gas producing a spiral-shaped cone as the
nucleus rotates.  When viewed from outside the cone, the spiral appears as an
oscillating jet centered on the spin axis.  Although this feature was seen
only near perihelion, the viewing geometry changed dramatically ($>40^\circ$)
during that time, introducing ample parallax for the use of intersecting
planes.

\begin{figure}
  \centering
    \includegraphics[width=85mm]{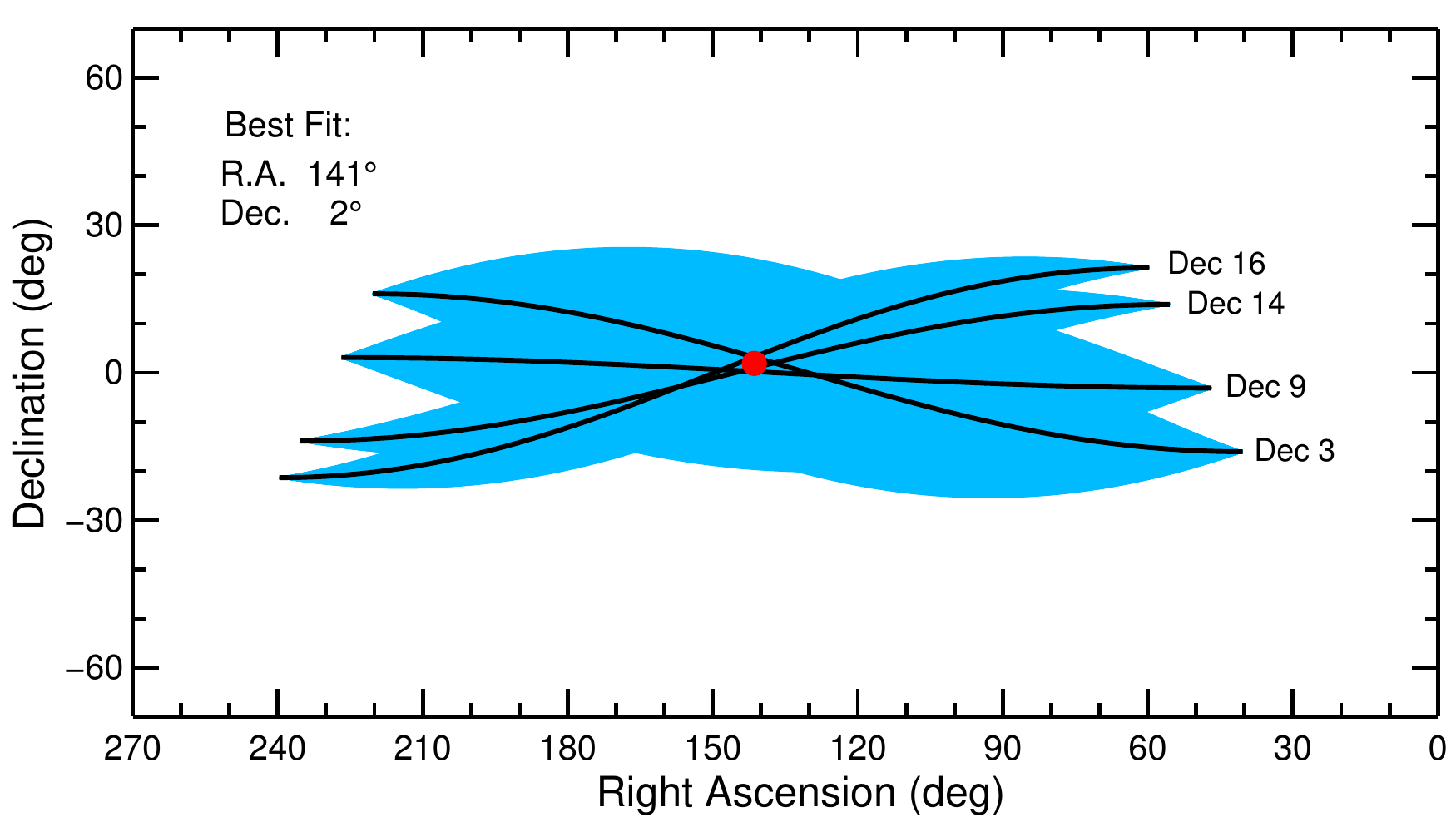}
  \caption{Pole solution showing the P.A./line-of-sight planes from the four
observation dates, projected as great circles onto the celestial sphere.  The
intersection of the planes defines the pole orientation in three dimensions.
The blue envelopes denote the uncertainties in each of the P.A. measurements.
The best fit solution is given on the figure; as explained in the text, the exact opposite pole direction is an equally viable solution and agrees with our Monte Carlo modeling.}
  \label{fig:gc_solution}
\end{figure}

We can discern the corkscrew in observations from December 3, 9, 14 and
16, and in each case, we estimate the center of oscillation to lie at a $\mathrm{P.A.}\sim90^\circ$.  
Because of the diffuse nature of the broad jets,
these are crude measurements with large uncertainties ($\pm$20$^\circ$ on December 4
and 9, and $\pm$10$^\circ$ on December 14 and 16).  Using these P.A.s, we compute a
best fit pole solution at $\mathrm{R.A., Decl.} = 141{^\circ}, +2^\circ$ with an uncertainty of
$\sim$15$^\circ$.  The sense of rotation cannot be determined from the
observations of the oscillating jet, so we cannot state whether this, or the
180$^\circ$ opposite solution (321$^\circ$, $-2^\circ$) is the positive pole.
Figure~\ref{fig:gc_solution} shows the projections of the derived planes onto the celestial
sphere, with their intersections defining the pole direction.  This solution
compares favorably to the result derived from the detailed modeling (if we
accept that the corkscrew defines the negative pole),
though the accuracy of the result, differing by only ${\sim}4^\circ$, is probably
somewhat fortuitous given the uncertainties in the measurements.

Having fixed our pole solution based on the modeling of the entire apparition, 
we could finalize the latitudes and the relative longitudes of the two source 
centers, along with their respective source radii. The primary, near-equatorial 
jet (shown in yellow in figures presented later in this section) is centered (to the nearest 5$^\circ$) at a 
latitude of $-$5$^\circ$ and has a radius $>$40$^\circ$ with a best estimate of $\sim$50$^\circ$. 
The secondary, ``southern'' jet (shown in green in figures presented later in this section) is centered at a latitude of $-$20$^\circ$ 
with a radius $>$35$^\circ$ and a best estimate of $\sim$40$^\circ$. The secondary source 
is centered at a longitude $\sim$160$^\circ$ greater than the primary source (with 
specific values tied to the sidereal period solution). Note that the center 
positions and the source sizes are somewhat interconnected as we needed to match 
either the presence or absence of material arising from the northernmost and southernmost 
parts of each source region. There is also some evidence that the secondary 
source is elongated approximately north-south, and may be somewhat narrower in 
longitude than in latitude. 
Also note that our modeling adopts a geocentric coordinate system, with
emission normal to a spherical surface.  Additional knowledge regarding the
shape of the nucleus would be required to project these results back to
their origins on a non-spherical topographic surface.

\begin{figure}
  \centering
    \includegraphics[width=85mm]{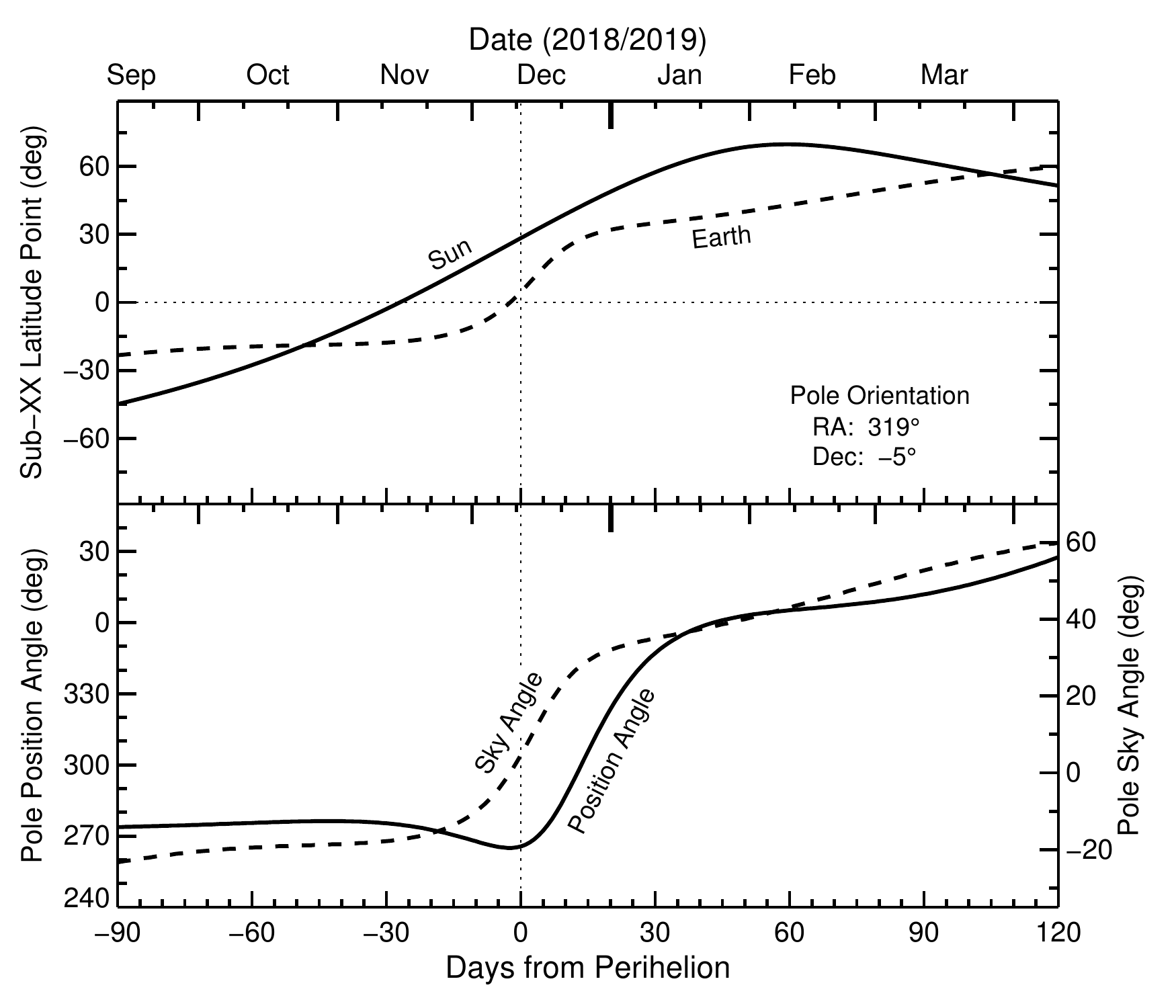}
  \caption{Sub-solar and sub-Earth latitudes plotted (top) as a function of time 
from perihelion, based on our preferred pole solution given in Table~\ref{t:pole_solution}
from modeling the CN images. Note the near-steady increase in the sub-solar latitude 
(solid) until late in the apparition, while the sub-Earth latitude 
(dashed) varies rapidly during the few weeks surrounding closest approach to Earth (December 16). 
The orientation of the positive (``northern'') pole is plotted (bottom) as a 
pair of curves, with the P.A. of the projected pole (solid) and the projection angle 
from the plane of the sky (dashed).}
  \label{fig:sub_lat}
\end{figure}

In addition to the requirements just discussed regarding source sizes and the need 
for reproducing features near the edges of the jets, another constraint came from 
the relative brightnesses of the two jets. In early November, they appeared nearly 
equally strong, while the primary jet dominates late in the apparition. As shown in 
the top panel of Figure~\ref{fig:sub_lat}, the sub-solar latitude varied substantially during 
the apparition, with a value of $-$49$^\circ$ only 100 days prior to perihelion, to $-$14$^\circ$ 
at the start of our imaging on November 1, to $+$28$^\circ$ at perihelion, 
$+$50$^\circ$ on January 3, and $+$70$^\circ$ at the end of our imaging on February 9. 
Thus the solar intensity favors the secondary source early on, but the southern portions 
of the secondary source would be continually in darkness late in the apparition. 
Based on these factors, in the final model, we assign 60\%\ of the particles to the 
primary source, and 40\%\ to the secondary source, for a 1.5:1.0 ratio, close to 
the simple ratio of areas of a 50$^\circ$ radius source as compared to a 40$^\circ$ radius region
(1.53).
These source sizes also correspond to about 18\%\ and 12\%\ of the total area of 
a spherical nucleus, significantly less than the nominal area required to match 
the water production rate, though the real
fraction of the surface could be significantly different depending on the
actual shape of the nucleus (see Section~\ref{sec:water_prod}). 
However, as already discussed, 
OH imaging suggests that at least some of the water must be released as icy grains, 
providing a natural explanation for the apparent discrepancy.

Due to the very broad nature of the jets, the measured outflow velocity in 
the plane of the sky is nearly the same as the actual radial velocity. This is 
also directly associated with the round, shell-like appearance of the outward moving 
features near perihelion, similar to the planetary nebula effect \citep{samarasinha00,schleicher04}. Ultimately, 
a near-constant parent outflow velocity of 0.76 km s$^{-1}$ works very well throughout the apparition. 
We did not detect a decrease with heliocentric distance, but the change in distance 
was relatively small during our imaging (from 1.19 to 1.06 to 1.31 AU), and the 
poorer signal-to-noise prevented accurate jet measurements at larger $r_\mathrm{H}$ to look for 
reduced velocities. Including an assumed heliocentric dependence would not affect the modeling or interpretation due to the small distance range. While our value of 0.76 km s$^{-1}$ is slightly smaller than calculations 
from early hydrodynamic models \citep[cf.][]{combi04}, it is in good agreement 
with more recent modeling of low production rate comets \citep{shou16}.

\vspace{-2mm}
\subsection{The Changing Sidereal Period}

\begin{figure}
  \centering
    \includegraphics[width=85mm]{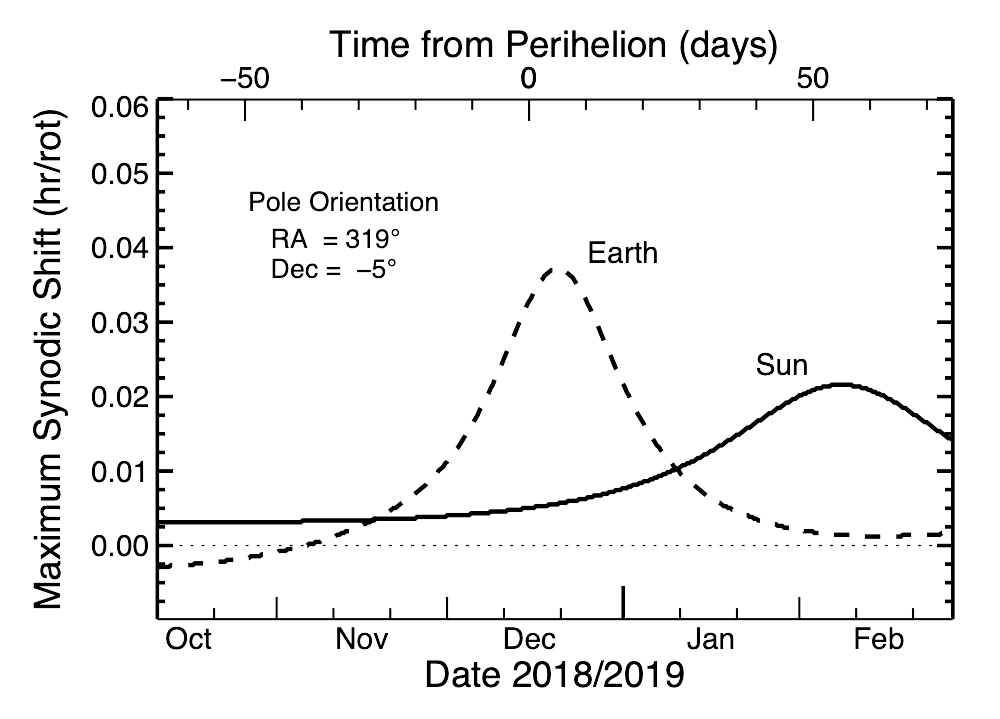}
  \caption{Nominal difference between the solar- (solid) and observer-based (dashed) components of the synodic, i.e., measured, period with the fundamental sidereal rotation period as a function of time around perihelion. The synodic shifts are calculated for a pole obliquity of 70$^\circ$, as derived from our preferred pole solution (Table~\ref{t:pole_solution}) from modeling of the CN images. The total offset between the apparent periods measured throughout the apparition are expected to be close to the combined values shown here, but other issues such  as changing projection effects can slightly modify this.}
  \label{fig:syn_shift}
\end{figure}

Because all model calculations use perihelion as the fiducial and time from perihelion 
to compute the rotational phase for any given observation, we must know the 
``accumulated phase'' throughout the apparition. While this is trivial when the 
rotation period is constant, it is very difficult when the period is changing in 
an unknown manner. A basic $4^{th}$ order fit was made to the apparent periods determined 
in Paper I, and this worked well as an estimate of the instantaneous apparent period 
at any time from early November to early February. However, two complications arise 
for our modeling. First, all model calculations require the sidereal rather than the 
synodic or apparent period. Using our pole solution, the offset between these reaches a maximum at
about $+$0.04 hr per rotation at closest approach to Earth (December 16) for the changing viewing 
geometry, while the offset due to the time from one local noon to the next on the 
comet reaches a maximum at about $+$0.02 hr per rotation approximately when the sub-solar latitude peaks 
(at $+$70$^\circ$) in early February (see Figure~\ref{fig:syn_shift}). How these affect the detailed jet 
morphology is quite complicated because projection effects are also varying throughout, 
but in all we expect the true sidereal period to match the apparent period in 
early November, to be $\sim$0.01--0.04 hr per rotation less near perihelion/closest approach to Earth in mid-December, 
and be between $\sim$0.00--0.02 hr per rotation less in early February. Thus, while the curve from 
Paper I provides an excellent guide as to the general changes in the sidereal period, 
one cannot simply apply an offset to this curve to obtain the correct curve to use 
for the sidereal period as a function of time. 

The second complication arises in that any error in the sidereal period as a 
function of time will accumulate as one is further in time from perihelion. 
As an example, even only a small 0.01 hr per rotation offset or error grows to an 
accumulated offset of 1.50 hr by February 8, corresponding to $\sim$60$^\circ$ longitude 
error for source region locations. Uncertainties associated with the detailed 
shape of the Paper I curve, especially near the extremes early and late in the 
apparition, coupled with the just noted estimated offsets between apparent and 
sidereal periods, makes it impractical to precisely determine the sidereal period 
for any particular observing run. Fortunately, we can partially mitigate this 
problem by forcing the source longitudes to remain fixed, but within the model 
this requires determining the average sidereal period for two dates -- perihelion 
and the particular date of observation. When distinct morphological features are 
evident, as in early November and in early January to the end of our imaging, 
precise values for this average can be determined, and were used when making the 
final refinements. However, since even the rate of change in period over any 
particular interval is not tightly constrained, the specific sidereal period for 
any date cannot be precisely known. In other words, the integrated 
changes in the period are well determined but not knowing the rate of change 
everywhere, the instantaneous sidereal periods are less-well constrained. 
As a consequence, absolute 
values for rotational phases are also unknown, and the phase values given for 
specific images are only appropriate in a relative sense within an observing run. 
With these caveats, along with the requirement that the source longitudes remain 
fixed, we were able to establish the absolute source locations. Specifically, 
the primary source is centered at 130$^\circ$ longitude, while the secondary source is 
centered at 290$^\circ$, each with uncertainties of less than 10$^\circ$, and where the fiducial 
of 0$^\circ$ longitude is defined as the anti-sunward direction at time of perihelion.

\subsection{The Overall Model Solution}
\label{sec:overall_model}

Our resulting, overall model solution is summarized as follows, based on our 
attempts to match the morphological features seen during full rotational cycles 
during a three-month interval. Basic parameters, including the pole orientation 
and source centers and sizes, are presented in Table~\ref{t:pole_solution}. 
Already noted in Section~\ref{sec:model_constraints} was the large change in both the sub-solar and 
sub-Earth latitudes during the apparition, shown in the top panel of Figure~\ref{fig:sub_lat}. 
The 3-D pole orientation of the nucleus, as seen from Earth, also varied drastically 
with time. To better understand the rotational motion projected on the sky 
plane, the bottom panel of Figure~\ref{fig:sub_lat} plots both the projected position angle of 
the positive (northern) pole, and the angle of the pole to the plane of the sky 
where $+$90$^\circ$ corresponds with the pole pointing directly towards Earth.

\renewcommand{\baselinestretch}{0.8}

\begin{deluxetable}{lrrr}
\tabletypesize{\scriptsize}
\tablecolumns{4}
\tablewidth{0pt} 
\setlength{\tabcolsep}{0.02in}
\tablecaption{Summary of Results for Wirtanen Model Parameters}
\label{t:pole_solution}
\tablehead{\colhead{Nucleus parameters:}
}
\startdata
Pole obliquity						&&&70$^\circ$ \\
\multicolumn{3}{l}{Pole orbital longitude\ (i.e. principal angle)} & 240$^\circ$ \\
Pole R.A. 		 					&&&319$^\circ$ \\
Pole Declination 					&&&$-$5$^\circ$ \\
\multicolumn{3}{l}{Maximum sidereal rotation period}	& 9.14 hr\\
\\
\hline
Source Regions:\\
\hline
			&Latitude	&Longitude		&Radius\\
\hline
Primary 		&$-$5$^\circ$		&130$^\circ$	&50$^\circ$ \\
Secondary	&$-$20$^\circ$		&290$^\circ$	&40$^\circ$\\
\enddata
\end{deluxetable}

\renewcommand{\baselinestretch}{1.0}

\begin{figure*}
  \centering
    \includegraphics[width=185mm]{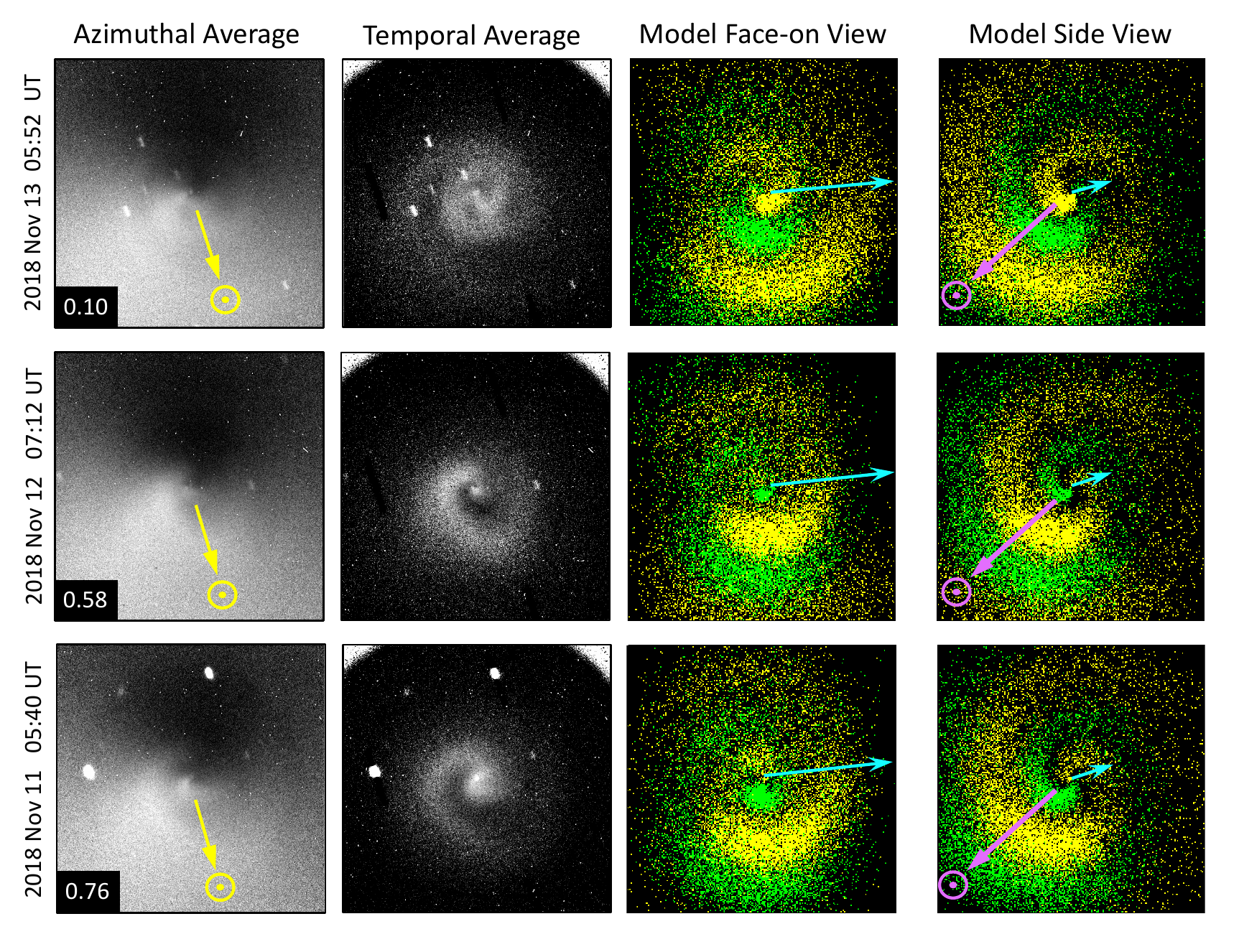}
  \caption{Representative images and corresponding model solutions from 2018 November 11--13. The first two columns contain enhanced CN images (enhancements given above the columns) with north up and east left, while the third column is as seen from Earth and the fourth column is as seen from the perpendicular direction from the right.
  All images on a row are for the time specified to the left of the first image. The rotational phase is given on the first panel of each row (see text for details). The direction to the Sun is indicated in yellow in the first column (valid for columns 1--3) and in magenta in the fourth column (valid for column 4). The turquoise arrow in the third and fourth panels indicates the projected direction of the north rotation pole (the differing lengths indicate the amount of projection for that view). The yellow and green points in the model images correspond to material emitted from the primary and secondary source regions, respectively, as further described in the text. All images are 60,000 km on a side and are centered on the nucleus.}
  \label{fig:model_images_nov}
\end{figure*}

\begin{figure*}
  \centering
    \includegraphics[width=185mm]{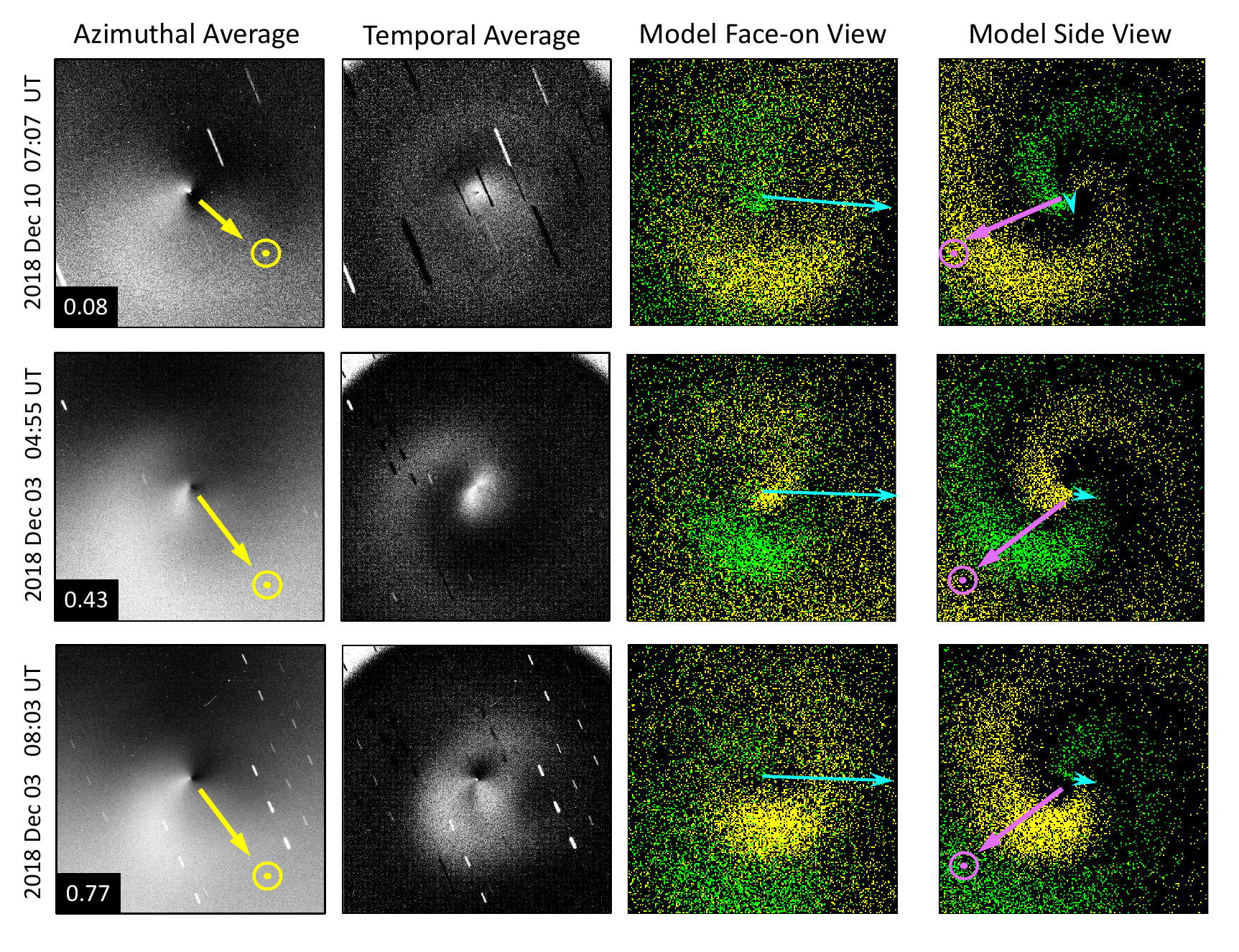}
  \caption{Representative images and corresponding model solutions from 2018 December 3--10. Images as described in Figure~\ref{fig:model_images_nov} except that the images are 30,000 km across.}
  \label{fig:model_images_dec1}
\end{figure*}

\begin{figure*}
  \centering
    \includegraphics[width=185mm]{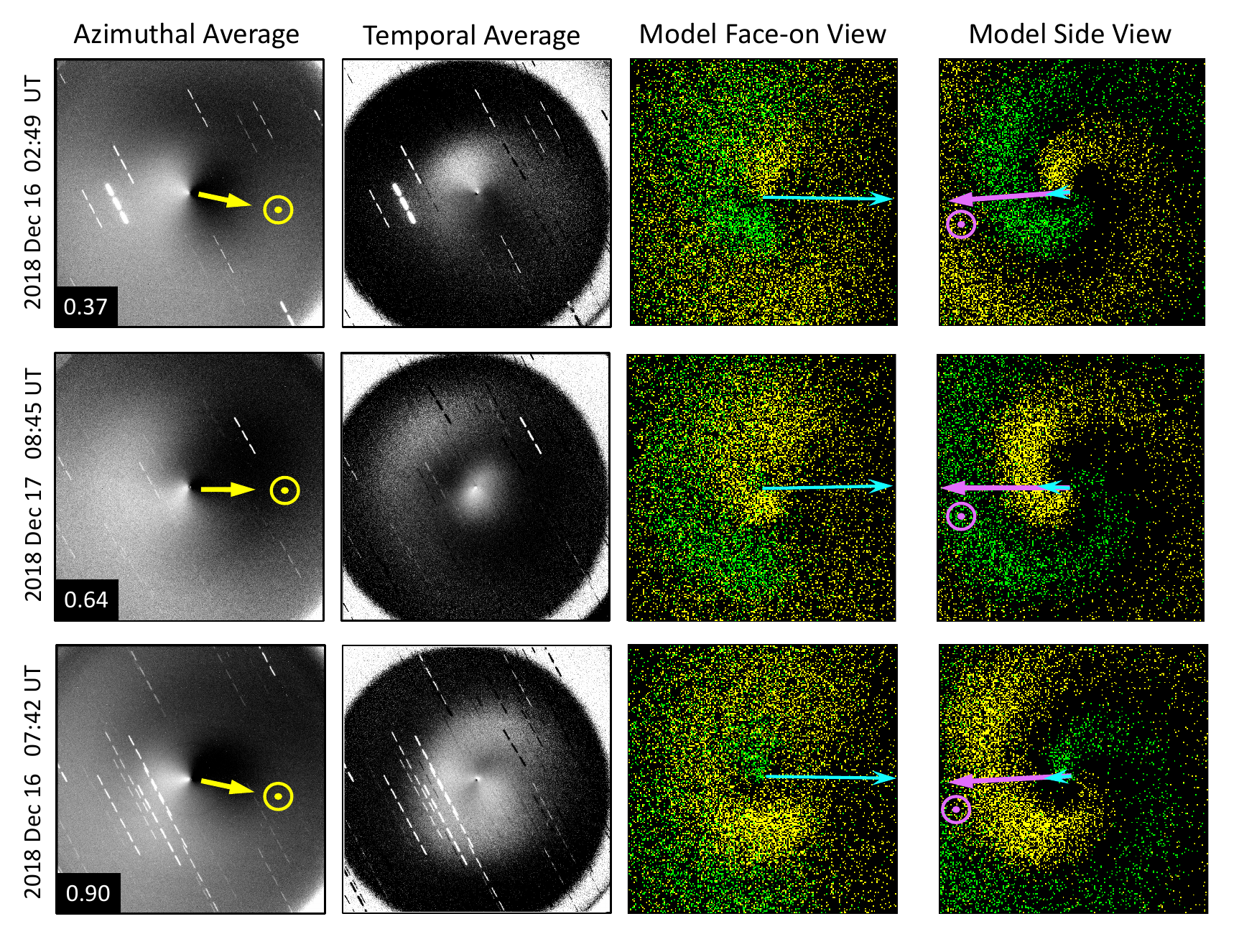}
  \caption{Representative images and corresponding model solutions from 2018 December 16--17. Images as described in Figure~\ref{fig:model_images_nov} except that the images are 30,000 km across.}
  \label{fig:model_images_dec2}
\end{figure*}

\begin{figure*}
  \centering
    \includegraphics[width=185mm]{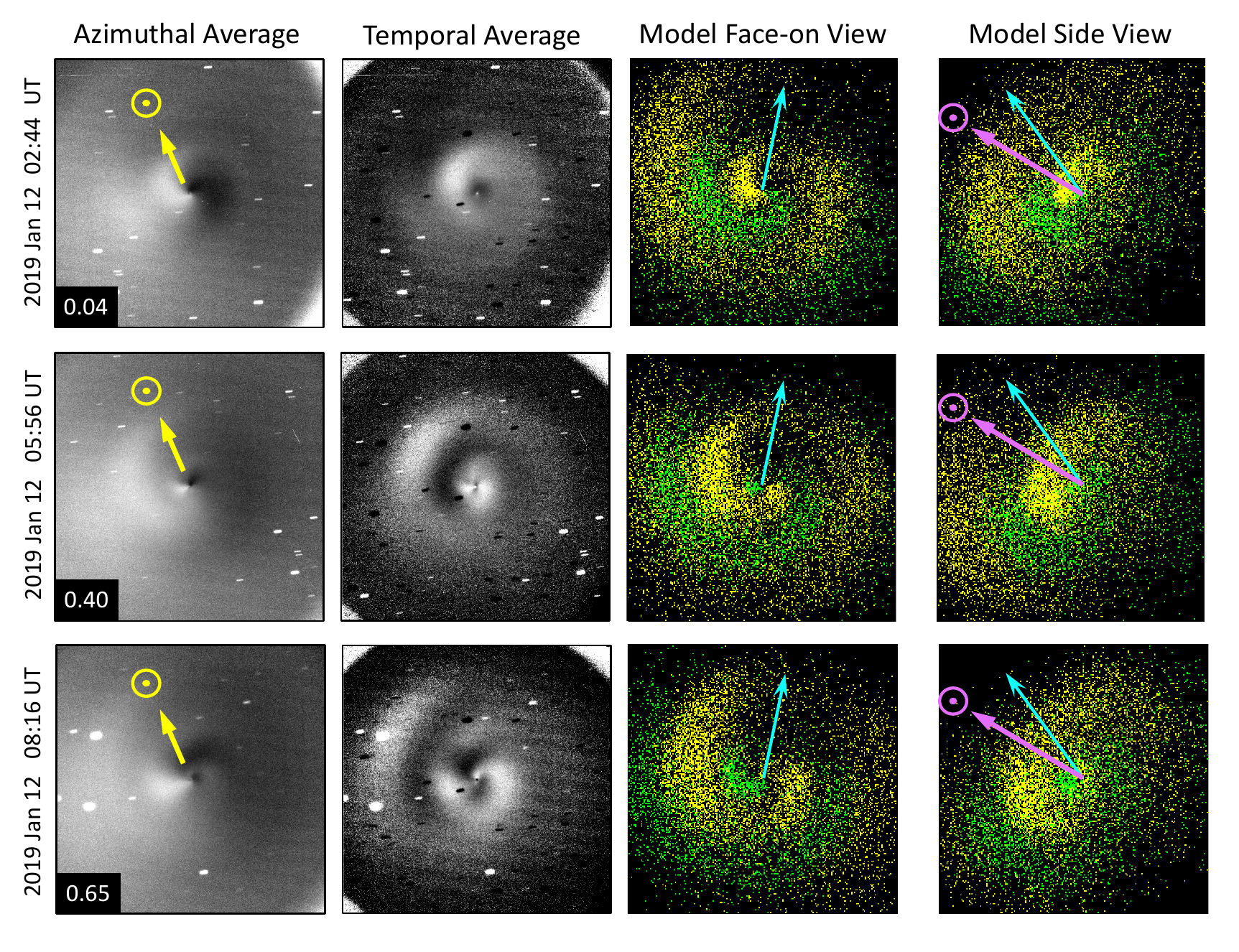}
  \caption{Representative images and corresponding model solutions from 2019 January 12. Images and scale as described in Figure~\ref{fig:model_images_nov}.}
  \label{fig:model_images_jan}
  \label{lastfig}
\end{figure*}

\vspace{-4.5mm}
Representative images, 
chosen to exhibit distinctive features throughout the apparition, are presented 
as a series of quadruple frames in Figures~\ref{fig:model_images_nov}--\ref{fig:model_images_jan}. 
In each case, 
we show the CN image, enhanced using azimuthal average (left) and using temporal 
average (center-left) along with the corresponding model solution (center-right). 
Additionally, we show a side-on view of the model (far-right) from the right side 
orthogonal direction to better understand the 3-D characteristics of the jets.
Because of the large width of each 
jet, a considerable portion of each rotational cycle has overlapping features as 
seen in projection on the sky. While these are not nearly as constraining as those 
with clear features, a few are shown for completeness. An advantage of the model 
is that we can color-code the jets, and we have assigned yellow to the primary 
jet, whose source is centered at $-$5$^\circ$ latitude, while we use green for particles 
originating from the secondary jet centered at $-$20$^\circ$. As already noted, in 
addition to the broad width of each jet simply due to the very large source 
radius, 50$^\circ$ and 40$^\circ$ respectively, there is significant additional dispersion 
due both to the daughter excess velocity and a Gaussian spread from the normal 
direction at each location within the source regions. 

Looking first in detail at the panels of Figure~\ref{fig:model_images_nov}, note that the features
in the top and bottom row of images are readily separated into the primary
(yellow) and secondary (green) jets but that the apparent single misshapen
spiral in the middle row temporal enhanced image is, in fact, the overlap of
the two jets, each dominating in different regions of position angle.
Also note the arrows showing
the projected direction of the Sun (yellow far-left and magenta far-right)
along with the projected rotation axis of the nucleus (cyan right and far-right).
While the jet morphology first appears as if one is viewing from nearly pole-on,
in fact the pole is nearly in the plane of the sky, and as previously discussed
the very broad jets produce this planetary nebula effect. The images enhanced with
the azimuthal average clearly are brightest towards the lower-left, consistent
with the phase lag of 1--2 hours with activity strongest in the local early afternoon;
this characteristic is not included in the model, which is therefore brightest
in the solar direction. 

As is evident by the projected pole orientation in the side-on model views in
Figures~\ref{fig:model_images_dec1} and \ref{fig:model_images_dec2}, the 
Earth crosses the equatorial plane only two days before
perihelion (see also Figure~\ref{fig:sub_lat}). With the center of the primary jet located at
$-$5{\degr}, nearly equal amounts of material are released to the north (left) and
south of our view on December 10 (Figure~\ref{fig:model_images_dec1}, top row) while the secondary jet
has a greater amount toward the southern hemisphere. In contrast, only a week
before (Figure~\ref{fig:model_images_dec1}, middle and bottom rows), the clockwise rotation is still
visible. We attribute some of the apparent mismatches between the model and the
images in the bottom row of Figures~\ref{fig:model_images_dec1} and \ref{fig:model_images_dec2} to our model assuming uniform
activity across the entire source region while any actual region subtending
$\sim$90\degr\ would naturally have pockets of higher or lower activity.
Finally, by January the overall morphology has a clear counter-clockwise rotation
(Figure~\ref{fig:model_images_jan}), and the model provides an excellent match throughout the rotational
cycle, often with what appears to be a single feature but is actually an overlap
of the two jets.

As is evident by simply comparing the two types of enhancements to the CN images, 
while the majority of the gas throughout the apparition is on the sunward 
side from the nucleus, as would be expected, a non-negligible amount of material 
is seen on the other side. While some of this anti-sunward material is simply 
due to projection effects, the model confirms that material must 
be continuing to be released from the nucleus even when the Sun is not 
illuminating the surface, as evidenced by the full spirals associated with the 
primary source. A first-cut result is that the CN parent is being 
released during the night at about 10\%\ of the rate as when the Sun is at the 
local zenith, and this value is included in the model. However, it is unknown 
whether this is due to a slow thermal lag of basic water ice vaporization or due 
to ongoing vaporization of CO or CO$_2$. Additionally, we see 
clear evidence for short-term thermal lag of about an hour. When closely 
examining the relative brightness of features as a function of position angle 
in the azimuthal enhancement, the true peak in release rates comes not 
when the Sun crosses the meridian, but up to an hour later. Indeed, we also see 
evidence for a lag at dusk, with more material released following sunset than 
at a corresponding time prior to sunrise. Since such short lags are not included 
in the model, we simply note these characteristics here but made no attempt 
to reproduce them.

As a relatively simple model, having a spherical nucleus with just two circular 
active regions, the model overall works remarkably well. In fact, there were 
only two instances when there were notable issues. The first was on December 12 
which, as detailed in Paper I, was during the aftermath of a significant outburst, 
and the ``standard'' morphology was overwhelmed by material released in the outburst. 
Once the cause of the discrepancy was identified, we stopped further model analysis. 
The second took place only a few days later (December 16 and 17) when a persistent 
feature was evident in the CN images very close to the nucleus at a P.A. of $\sim$60--90$^\circ$, 
and that apparently did not change in appearance with rotation. Examination of 
the continuum images confirmed that this was a dust feature contaminating the 
CN filter images. Note that the timing was just days prior to the comet reaching 
minimum phase angle and just following the Earth's crossing of Wirtanen's 
orbital plane. Thus a tail would have had the greatest projection effect and would 
appear both shorter and brighter than at other times. 
It is also possible that larger, very slow moving grains released in the 
December 12 outburst might have finally moved beyond the seeing disk. 
Whatever the cause, and having failed to reproduce this feature with the 
existing model (even by attempting to add a third, small source region), we 
ultimately ignored this stationary inner-coma feature.
Finally, note that the second outburst reported in Paper I, that was evident 
in $R$-band images on January 28, was smaller than the December 12 outburst and not detected in the CN morphology.

Other, ongoing differences between the images and the model we attribute to 
limitations of the model. We are certain that Wirtanen's nucleus is not 
spherical in shape, and also that local topography will play a role. Source regions 
won't simply be circular, and are unlikely to be uniformly active over their 
entire region. Most surprising is how regular the overall activity is 
every rotational cycle, perhaps because of the large size of the source regions: 
as ice is vaporized away in some sub-regions, more ice is exposed in other areas 
but all within the same broad feature. We also acknowledge that, based on 
spacecraft imaging of several comets, there are undoubtably other, smaller 
active regions distributed across the remainder of the nucleus' surface, but that 
the majority of the coma material arises from the two large source regions that 
we have identified here. 

Our model assumes simple, i.e., principal axis, rotation. As discussed in Paper I, the
CN morphology repeated very well from cycle-to-cycle and night-to-night suggesting
this was a valid assumption. We did not see evidence for systematic or repeating 
deviations in CN morphology compared to our model which might indicate a non-principal 
axis component to the rotation. Thus, we conclude that our assumption of simple rotation
is safe and that any non-principal axis rotation is small or non-existent.

\section{DISCUSSION}
\label{sec:discussion}

\subsection{Synergistic aspects of imaging, modeling, and photometry}

The recent, excellent apparition of Comet 46P/Wirtanen provided an unusual 
opportunity to obtain both narrowband photometry and imaging over a wide range 
of distances and viewing geometries. Each method of investigation provided 
interesting and sometimes unexpected results, but when combined with a successful 
model of the gas jet morphology, a complete and self-consistent characterization 
of Wirtanen's nucleus is revealed. 

For instance, measured production rates of the carbon-bearing species (CN, C$_2$, and C$_3$) 
exhibit little to no asymmetry before and after perihelion. Not only is this lack of 
a pre-/post-perihelion asymmetry atypical of Jupiter family comets \citep{ahearn95}, 
it would generally imply that the nucleus had a low obliquity and thus minimal seasonal 
variations. However, the model solution yielded a quite large obliquity (70$^\circ$), 
and only the fact that the major source regions were located very close to the equator 
prevented the correspondingly large seasonal effects. Examination of the images 
obtained near closest approach to Earth revealed another puzzle: the distinct spiral 
structures observed in early November were replaced by an overall top-to-bottom 
motion of fairly diffuse features. Only with the modeling did it become clear that 
the jets were extremely broad and that we (at Earth) took many weeks to traverse 
across the jets as they swept by the Earth each rotational cycle. This also meant 
a continually changing overlap pattern of the two jets, adding to the complexity 
of the observed morphology. Ultimately the 
spiral pattern returned in January, but with the opposite sense of rotation due to 
our then viewing the comet from the other hemisphere, having completed our crossing 
of the equatorial plane. 

Other key findings include the discovery of two major source regions, both extending 
across the equator but at nearly opposite longitudes. This explains the similar but 
yet distinct appearance of alternating spiral structures seen in the CN gas images 
early and late in the apparition. The spacing of the alternated jets was nearly the 
same but one jet had complete spirals while the other one's were either incomplete 
or swept by the Earth and were therefore difficult to detect. Ultimately, 
modeling revealed that the same two features (and thus the same source regions) 
dominated activity throughout the apparition, but that the smaller and more southerly 
located source region was proportionally less important late in the apparition when 
the sub-solar latitude was far to the north. However, since no pre-/post-perihelion 
production rate asymmetry is evident for the carbon-bearing species, the southern 
source's center at $-20^\circ$ coupled with the Sun being directly overhead in October 
compensated for the Sun crossing the equator nearly a month prior to perihelion. 
The longitudinal separation of the sources (160\degr) also naturally explains our finding from Paper I that the lightcurve peaks were separated by an effective longitude of $\sim$170\degr.

\subsection{Inter-species comparisons}

Unlike the carbon-bearing species, that show clear jet structures in their 
morphology and no production rate asymmetries about perihelion, the hydrogen-bearing 
species (OH and NH) are significantly different in both respects. 
The {\rh}-dependencies of OH and NH are only slightly steeper than the carbon species 
prior to perihelion with log-log slopes near $-$4.3, but were both significantly 
steeper (slopes near $-$5.1) outbound. If one uses water (i.e., OH) as the fiducial, 
then the simple explanation is that there is a difference in the abundance ratio 
of the carbon species between the two source regions with the higher abundance 
of carbon in the primary (equatorial) source region. 

Another difference between hydrogen-bearing and carbon-bearing species is that 
OH and NH show little to no evidence of jet morphology that is clear in CN and 
the carbon-chain species. Moreover, there is a bulk enhancement in the anti-sunward 
hemisphere, indicating that a significant portion of OH and NH arise from icy grains. 
This tailward asymmetry is not, however, equally distributed on either side of the 
dust tail, but is enhanced towards the peak in brightness of the carbon-bearing 
jets, i.e., it appears that icy grains are released from the same source regions as 
create the CN jets, but these grains are then pushed in the anti-solar direction by 
solar radiation pressure. Because no motion of the bulk OH morphology is evident 
during a rotation cycle, nor is jet-like structure seen, we think that the grains 
must be relatively large and slow moving. Also, having an icy grain source would 
effectively make OH a granddaughter species, further dispersing the OH. 
Why these icy grains contain water and ammonia but not the carbon-bearing species 
remains a mystery, but Wirtanen is only the latest of several comets displaying 
such behavior, including 103P/Hartley 2 \citep{knight13a}, C/2013 R1 Lovejoy \citep{opitom15b}, and 252P/LINEAR \citep{knight16b}
Unfortunately, we were unable to assess if the proportion of OH and NH arising from icy 
grains was related to the steeper {\rh}-dependencies measured after perihelion, or if the 
proportions of icy grains arising from the two source regions are similar. The lack
of rotational variation in OH and NH morphology and the limited number of narrowband 
images  prevented us from assessing if the relative abundances in the two
source regions varied with time. It is unlikely that
the flatter pre-perihelion OH and NH {\rh}-dependencies was due to to the presence of
a large population of icy grains released significantly before perihelion since, at these heliocentric distances, the lifetime
of such grains should be relatively short \citep[hours to days;][]{beer06} compared to the
temporal span of our observations.

The primary attributes of the dust coma are also interesting. A clear dust tail 
is seen but the background dust coma is brighter towards the peak in CN 
brightness, similar to that for OH. Dust jets are not detected, but we attribute 
this lack of discrete features as being due to much lower outflow velocities coupled with 
a large dispersion in particle sizes and the associated large dispersion in velocities. 
This combination would effectively wipe out any features. Somewhat of a surprise is 
that the dust abundance, as measured by {\afrho} and following adjustments for 
phase angle and aperture size, has an {\rh}-dependence the same 
on either side of perihelion, i.e., matching the carbon-bearing species. 
This raises the intriguing possibility that H$_2$O may not be the primary driver
of Wirtanen's activity, or at least that another volatile parent species contributes
non-negligibly. However, Wirtanen was found to be poor in CO \citep{biver19a,saki20}, while 
derived atomic production rates in the far UV suggest that CO$_2$ and O$_2$ were ``not
abundant'' in comparison to H$_2$O \citep{noonan21}. Thus, there is no obvious
volatile species that might explain the behavior.

\subsection{Long-term Secular Decrease in Production Rates}

Comet Wirtanen exhibited a strong decrease in both gas and dust production rates 
since the 1990s. In fact, our data show a monotonic 
decrease from 1991 to 1997, then to 2008 and now to the 2018 apparition. Due to 
the comet's greatly differing distances from Earth, and the corresponding projected 
aperture sizes, we looked for aperture effects associated with our Haser model 
parameters for each gas species and eliminated this as a cause. There was, as 
with most comets, a significant aperture effect for {\afrho}, for which we measured and 
corrected. Also adjusting for phase angle effects, {\afrhoz} showed a very 
similar amount of secular decrease as the gas species: just over a factor of two. 
Our results are very similar to that reported by \citet{combi21} from 1997 
to 2018, who also claimed a factor of two near perihelion and an even larger 
decrease at larger distances. Combi et al.\ also found that Wirtanen had substantially 
higher values for most of 2002 than in 1997, but large variations in only a few days 
on multiple occasions strongly indicate that most of their 2002 observations were 
dominated by outburst activity. Thus, the ensemble evidence suggests that the underlying 
activity level (sans outbursts) indeed has been continuing to decrease for the 
past 20--30 years.

This rate of drop in production rates, while unusual, is not unprecedented. Of the 
18 comets in our photometry database for which we have multi-apparition data suitable 
for investigating secular changes, only a handful exhibit clear changes not 
attributable to a change in perihelion distance. In particular, Comet 9P/Tempel 1 
had decreases by factors of 1.5 to 2.4 depending upon species between 1983 and 2005 
\citep{schleicher07a}, while the most extreme case is 103P/Hartley 2, where all 
species decreased by a factor of $\sim$3 from 1991 to 2010/11 \citep{knight13a}. 
One common attribute of Hartley 2 and Wirtanen is that each had fairly recently 
undergone a major orbital change from significantly larger perihelia due to close 
passages with Jupiter. Hartley 2 suffered a large perturbation in 1971, with a 
decrease in perihelion from 1.67 AU to 0.90 AU and had crept back to 1.06 AU by 2010.
Wirtanen's perihelion decreased from 1.61 AU to 1.26 AU in 1972, followed by another drop to 
1.08 AU in 1984 (it is now at 1.05 AU). The other notable attribute is that both 
objects were classified as ``hypervolatile'' based on their water production rates 
being higher than could be understood by vaporization of their small nucleus surface 
areas. Indeed, both exhibit strong evidence of icy grains, explaining the hyperactivity. 
What is unclear is how these two characteristics relate to one another. Are icy grains 
associated with the decrease in perihelia and the resulting higher surface temperatures, 
i.e., are icy grains preferentially released in these circumstances? Also, are 
the decreasing total production rates associated with a decrease in the amount of 
icy grains? The latter possibility appears to {\it not} be the case, as the icy 
grains apparently do not include carbon-bearing species and yet all species show 
the same secular decrease.

\subsection{Fractional Active Areas of the Jet Source Regions}

Based on our modeling of the two jets throughout the 2018/19 apparition, we
concluded that the primary source region had a radius of $\sim$50\degr\
while the secondary source was also quite large, having a radius of
$\sim$40\degr\ (18\%\ and 12\%\ of the total surface area, respectively).  This
is near the high-end for the sizes of jets derived from coma models, but it
is not anomalous.  Comets C/1995~O1 Hale-Bopp and 67P/Churyumov-Gerasimenko exhibited
jets of radii 15--30\degr\ \citep{vasundhara99,schleicher04b,schleicher06b},
and C/2007~N3 Lulin had jets of 10\degr\ and 20\degr\ \citep{bair18}, but
the most comparable comet previously modeled was C/1996 B2 Hyakutake
\citep{schleicher03a}, which had a primary source of $\sim$56\degr\
radius (and a smaller secondary source of $\sim$16\degr).

In 2018, Wirtanen's measured active fraction was 55\%.  This suggests that
the 30\%\ of the surface covered by the two jet sources represents a
significant portion of the comet's activity if we assume that the source of
the CN also reflects the source of water. Indeed, given that independent
evidence suggests that part of the water is emitted from grains in the coma, 
it is likely that the two jet sources represent the
majority of the comet's activity.  The extent of these active areas,
distributed over a non-spherical nucleus increases the likelihood that 
ejected material will produce reactionary forces that do not pass through the
nucleus' center of mass, resulting in torques that can alter the comet's
rotation. This is discussed further in the next subsection.

Regardless of whether we assume that the active fraction is 55\% based on 
water production rate or 30\% based on CN jet modeling, it is evident that Wirtanen 
is still a ``young'' comet having ample volatile ices near the surface. Assuming that active fraction decreases as a comet spends
more time in the inner solar system due to loss of accessible volatiles and/or 
mantling over of the surface, a larger active fraction should be indicative
of a younger comet. Of 38 Jupiter family comets 
having measured nucleus sizes in our photometric database \citep{schleicher16b}, 
27 have smaller active fractions than 30\% (our more conservative number), 
and two-thirds of these are smaller than 10\%. 
Our database also reveals that active fractions of Jupiter family comets generally 
decrease with decreasing perihelion distance. Wirtanen's active fraction is high 
compared to the subset with comparable perihelion distances, further 
evidence that it has not spent as much time in the inner solar system.
We note, however, that this assessment of ``youth'' greatly simplifies a host of potentially competing factors including compositional differences and unknown dynamical histories, and competing interpretations exist \citep[e.g.,][]{belton10,vincent17}.

\subsection{The Rotation Spin-down and Spin-up}

A major result from our studies was the change documented in the rotation
period of Wirtanen, and the nature of that change in light of the modeling
results.  Before perihelion, the rotation period increased by $\sim$4\%\ \citep{farnham21},
only to decrease again after perihelion to a value similar to what it was at
the start of the apparition, indicating that the net torques on the nucleus
are significant, yet they are highly seasonal and largely cancel out when
integrated over the perihelion timeframe.  This is likely due to the sizes
and relative locations of the two active areas, which produce net torques
that change as the sub-solar latitude shifts north.  

One scenario that would explain the observed changes involves a prolate
spheroid nucleus rotating about its small axis (consistent with the radar estimate of the nucleus being 
$1400{\times}1100$~m; E.~Howell, priv.\ comm.), with the two jets placed at
angles $\sim$10\degr\ from each end, where the moment arms are long.  Because
the jets are effectively 160\degr\ apart in longitude, they are offset on the
same side of the body, producing torques in opposite directions.  When the
Sun is in the southern hemisphere, the smaller jet receives more direct
insolation and thus dominates the torques during the pre-perihelion
timeframe, slowing the nucleus' spin.  As the Sun moves north, the southern
jet weakens and the equatorial jet, producing torques in the opposite
direction, becomes dominant, spinning the comet up again.  Although this
scenario qualitatively explains the observations, it is a simplistic
depiction and is by no means unique.

The only other comet known to exhibit a rotational change in which the period
reverses direction in the middle of an apparition is comet
67P/Churyumov-Gerasimenko, which spun down slightly during its approach to
the Sun before spinning up more dramatically around perihelion \citep{keller15, kramer19}.  
However, there are significant differences between
Wirtanen and 67P, in that 67P did not exhibit significant isolated jet
activity (the rotation rates were detected from {\it Rosetta} observations of the
nucleus, as opposed to lightcurves or coma morphology)
nor did 67P show the degree of volatile release from icy grains exhibited by 103P \citep{ahearn11a,protopapa14} or (based on our study) Wirtanen \citep[e.g.,][]{gasc17,biver19b,protopapa20}.  Thus, the torques on
the nucleus arose from the integrated activity across the non-spherical
nucleus, and the extreme seasonal differences between the northern and
southern hemispheres (active during the approach and perihelion timeframes,
respectively) resulted in the rotation reversal.

It is difficult to assess how common this phenomenon might be since few comets are observed
well enough over a sufficient baseline. Rotation periods are easiest to determine 
for comets with weakly active or inactive nuclei, where photometry can measure the light
reflected from the rotating nucleus directly. This is possible either when comets
are far from the Sun and their activity has decreased or at any time for intrinsically low 
activity comets. In both cases, the weak activity cannot provide the  
torquing needed to change the rotation period except under highly favorable scenarios
and, in the distant observation case, the geometry change is likely to be very slow. Thus, morphological studies or inner coma lightcurves are the most promising ways to detect changing rotation periods over short timescales. However, they require bright comets with reasonably small geocentric distances, favorable morphology, and/or short rotation periods. Over the last decade, only two other comets have yielded similar coverage of the rotation period as Wirtanen: 103P/Hartley 2 \citep[e.g.,][]{knight11b,samarasinha11,belton13a,knight15b} and 41P/Tuttle-Giacobini-Kres\'ak \citep{bodewits18,schleicher19}. In both cases, much larger rotation period changes were detected than for Wirtanen, but the change was always in the same direction. Neither comet has yet been modeled with a fully self-consistent solution that reproduces the coma morphology, but their monotonic changes in period suggest that neither has a scenario similar to Wirtanen.

Without several epochs of extensive observations in a single apparition, it is impossible to detect a small and reversing change in the rotation such as we measured for Wirtanen. The
increase/decrease in the rotation period raises an interesting
question: How much did the rotation state change and how does this factor
into comparative studies with other comets?  If we consider the start and end
points for the rotation period found in Paper I (e.g., the change per orbit), then Wirtanen
would be considered to exhibit little rotational variability. On the other
hand, we observe an increase and then a decrease totaling $\sim$21~min ($\sim$4\% of the rotation period),
within a few months of perihelion, indicating that there are significant
variations occurring over time.  Other comets could exhibit similar
tendencies, with seasonal effects producing opposite trends that cancel out
over the course of the orbit. Thus, objects known to have monotonic changes from apparition-to-apparition or even no change might be experiencing similar behavior to Wirtanen. 
In fact, this is exactly the case for 67P which, in the absence of {\it Rosetta} data, would be judged to have a monotonic period change. Without comprehensive models of pole orientation and jet location and extent, it is difficult to determine the cause of these changes. 

Another result of our modeling is how Wirtanen challenges predictions of changes in rotation period that are based on behavior being globally similar between comets. As just discussed, Wirtanen's change in rotation state is quite different from that of Hartley~2, a superficially similar object (see discussion in Paper I). The model of \citet{samarasinha13} and \citet{mueller18} uses overall activity and nucleus size to predict changes in rotation period. Their methodology predicts a period change of 18 min for Wirtanen. We measured a $+$9 min then a $-$12 min change, so either it is a very good prediction (for a total change of 21 min), off by $\sim$50\% (for either the spin-down or spin-up considered in isolation), or substantially off (for the $\sim$2 min net change over the apparition). \citet{rafikov18} predicted changes in the rotation periods of comets based on measured non-gravitational accelerations. Given Wirtanen's relatively large rate of period change over small segments of its orbit, it would be viewed as very prone to spin evolution if only observed during these portions, while its spin evolution would be viewed as relatively modest when considered on an orbit-by-orbit basis. 
An investigation of Wirtanen's non-gravitational forces during the subsets of time in which we saw the largest period changes would be an interesting test of the Rafikov model, though it is beyond the scope of this paper.

\section{Conclusions}
\label{sec:conclusion}

We successfully conducted a multi-faceted observing campaign of Comet 46P/Wirtanen 
during its historically close approach to Earth in 2018/19. 
These data, combined with photometric observations from three prior apparitions, and 
detailed modeling of the recent imaging, permitted us to obtain a much more complete 
understanding of this ``hyperactive'' comet than previously had been obtained despite it 
having been the original target of ESA's {\it Rosetta} mission. 

Paper I \citep{farnham21} presented results from our extensive CN imaging, finding a continuously changing rotation period that first increased then decreased. The interconnected studies presented here provide a means to explain and interpret the findings in Paper I. These include:
\begin{itemize}
\item Gas production rates are symmetric about perihelion for carbon-bearing species (CN, C$_3$, and C$_2$), but fall off more steeply post-perihelion for OH and NH. Despite the different trends, the composition remained ``typical'' throughout our observations.

\item There is an ongoing secular decrease in gas activity and dust production over four apparitions: 1991, 1997, 2008, 2018 (we did not obtain data in 2002 or 2013).

\item The carbon-bearing species exhibit similar coma morphology that varies with rotation. The coma morphology of OH, NH, and dust did not change appreciably with rotation. We attribute the appearance of OH and NH to their having a significant source from icy grains in the coma, while the lack of rotational signal of these and the dust is due to the lower velocities and larger velocity dispersion among grains as compared to gas.

\item Our preferred model replicates the cycle-to-cycle variation of the two CN jets identified in Paper I as well as the evolution over the apparition of their senses of rotation and relative brightnesses. The model has two large, mid-latitude jets and a pole oriented at $\mathrm{R.A.}=319^\circ$, $\mathrm{Declination}=-5^\circ$, based on an obliquity of 70\degr\ and a principal angle of 240\degr. 

\item The primary jet has a radius of 50\degr\ and is centered at a latitude and longitude of ($-5^\circ$, $130^\circ$), while the secondary jet is 40\degr\ in radius and is at ($-20^\circ$, $290^\circ$). The seasonal variation in production rates can, therefore, be explained if there is compositional heterogeneity between the source regions, with the northern region having a higher abundance of carbon.

\item The fractional active area in 2018/19 is $\sim$55\%. Our modeling suggests the source regions cover $\sim$30\% of the surface, implying that they represent the plurality of surface activity. The imaging suggests that much of the excess OH signal (above 30\%) may come from icy grains in the coma. Regardless of which measure is used to assess active area, Wirtanen is at the high end of our photometry database and appears to be a ``young'' surface.

\item Our model demonstrates that the changing rotation period discovered in Paper I represents a change in the sidereal period and is not due to changing viewing geometry alone. It also provides a natural explanation for the alternately increasing then decreasing sidereal rotation period as being due to the evolving dominances of the two source regions as their seasons change during the apparition. Such behavior is difficult to diagnose without extensive observations and, if common, likely skews our understanding of the stability of comet rotation periods.

\end{itemize}

Our findings, in concert with the diverse results obtained by the community over the same time period are a reminder that, for highly favorable apparitions such as Wirtanen, major new insights into and significant constraints on physical properties of comets can be obtained at a fraction of the cost of a spacecraft. Nonetheless, space missions are unrivaled in providing specific details which cannot be obtained remotely. Due to its orbital characteristics, Wirtanen has long been a prime mission target and continues to be proposed for a future spacecraft mission; our many findings should prove very useful in planning such missions.

\begin{acknowledgments}
We gratefully acknowledge Lori Feaga, Dennis Bodewits, Josie Schindler, Brian Skiff, Uwe Konopka, Carrie Holt, and Larry Wasserman for assisting with some of the observations, and Allison Bair for assisting with the photometric analyses and construction of tables.  We thank the anonymous referees for prompt and helpful reviews. This work was supported by NASA Solar System Observations Program grants 80NSSC18K0856 and 80NSSC18K1007, Hubble Space Telescope grant HST-GO-15372, and NSF award AST1852589. 
These results made use of the Lowell Discovery Telescope (LDT) at Lowell Observatory. Lowell is a private, non-profit institution dedicated to astrophysical research and public appreciation of astronomy and operates the LDT in partnership with Boston University, the University of Maryland, the University of Toledo, Northern Arizona University and Yale University. The Large Monolithic Imager was built by Lowell Observatory using funds provided by the National Science Foundation (AST-1005313).

\end{acknowledgments}

%

\facilities{Lowell Discovery Telescope, Lowell Observatory 42 inch (1.1 m) Hall Telescope, 
Lowell Observatory 31 inch (0.8 m) telescope.}


\software{IDL}
\clearpage

\bibliographystyle{aasjournal}




\end{document}